\documentclass[twocolumn]{aastex61}
\usepackage{comment}
\usepackage{color}

\newcommand{\henb}[2]{\frac{\partial #1}{\partial #2}}
\newcommand{\diff}[2]{\frac{\mathrm{d} #1}{\mathrm{d} #2}}
\newcommand{\BF}[1]{{\bf #1}}
\newcommand{\average}[1]{\ensuremath{\langle#1\rangle} }

\submitjournal{ApJ}

\shorttitle{Parametric Study of the RWI in a 2D Barotropic Disk II}
\shortauthors{Ono et al.}
\begin{document}

\title{Parametric Study of the Rossby Wave Instability in a Two-dimensional Barotropic Disk II: Non-Linear Calculations}

\correspondingauthor{Tomohiro Ono}
\email{ono.t@osaka-astro.jp}

\author{Tomohiro Ono}
\affil{Department of Earth and Space Science, Osaka University, Toyonaka, Osaka, 560-0043, Japan}

\author{Takayuki Muto}
\affil{Division of Liberal Arts, Kogakuin University, 1-24-2 Nishi-Shinjuku, Shinjuku-ku, Tokyo 163-8677, Japan}

\author{Kengo Tomida}
\affil{Department of Earth and Space Science, Osaka University, Toyonaka, Osaka, 560-0043, Japan}

\author{Zhaohuan Zhu}
\affil{Department of Physics and Astronomy, University of Nevada, Las Vegas, 4505 S. Maryland Pkwy, Las Vegas, NV, 89154, United States}



\begin{abstract}

Vortices in protoplanetary disks have attracted attention since the discovery of lopsided structures. One of the possible mechanisms for producing vortices is the Rossby Wave Instability (RWI). In our previous work, we have performed detailed linear stability analyses of the RWI with various initial conditions. In this paper, we perform numerical simulations of the vortex formation by the RWI in 2D barotropic disks using the Athena++ code. As initial conditions, we consider axisymmetric disks with a Gaussian surface density bump of various contrasts and half-widths. Perturbations grow as expected from the linear stability analyses in the linear and weakly non-linear regimes. After the saturation, multiple vortices are formed in accordance with the most unstable azimuthal mode and coalesce one after another. In the end, only one quasi-stationary vortex (the RWI vortex) remains, which migrates inward. During the RWI evolution, the axisymmetric component approaches the stable configuration. We find that the axisymmetric component reaches the marginally stable state for the most unstable azimuthal mode at the saturation and the marginally stable state for the $m\ =\ 1$ mode at the final vortex merger. We investigate the structure and evolution of the RWI vortices. We obtain some empirical relations between the properties of the RWI vortices and the initial conditions. Using tracer particle analyses, we find that the RWI vortex can be considered as a physical entity like a large fluid particle. Our results provide a solid theoretical ground for quantitative interpretation of the observed lopsided structures in protoplanetary disks. 
%check 2
\end{abstract}




\section{Introduction} \label{sec:intro}

Recent observations have revealed protoplanetary disks with lopsided structures especially in transitional disks \citep[e.g.,][]{2013Sci...340.1199V, 2013PASJ...65L..14F, 2015ApJ...812..126C}. 
It is important to know how the lopsided structures are formed and how they are related to the disk evolution and the planet formation. 
One possible mechanism for producing such lopsided structures is capturing of dust particles ($\sim$~mm in size) by a large-scale gas vortex. 
Theoretically, it has been known that a large-scale vortex of gas can survive for a long time \citep{2000ApJ...537..396G} and can efficiently trap dust particles due to gas drag \citep[e.g.,][]{1995A&A...295L...1B}. 
With a vortex induced by an embedded planet, \citet{2014ApJ...795...53Z} showed that some ALMA observations can be reproduced with three-dimensional (3D) MHD simulations including dust particles. 
%check 2

Several vortex formation mechanisms have been proposed; the Rossby wave instability (RWI) \citep[e.g.,][]{1978ApJ...221...51L, 1999ApJ...513..805L}, the baroclinic instability \citep[e.g.,][]{2003ApJ...582..869K}, the vertical shear instability \citep[e.g.,][]{1967ApJ...150..571G, 2013MNRAS.435.2610N}, the zombie vortex instability \citep[e.g.,][]{2013PhRvL.111h4501M}, and the pebble accretion onto planets \citep{2017MNRAS.467.3379O}. 
As far as the observations show, all the protoplanetary disks with the lopsided structures are transitional disks, which have an inner cavity. 
In the case of a transitional disk, the existence of sharp variations of physical quantities (e.g., surface density) is naturally expected. 
When a protoplanetary disk has the sharp radial variations, a large-scale vortex of gas is expected to be formed by the RWI. 
Therefore, the RWI is one of the most promising mechanisms for explaining the observed lopsided structures. 
%check 2

The RWI has been studied with linear stability analyses \citep{1999ApJ...513..805L, 2000ApJ...533.1023L, 2010A&A...521A..25U, 2013ApJ...765...84L, 2012MNRAS.422.2399M}. 
The existence of a vortensity local minimum is necessary for the RWI to take place \citep{1999ApJ...513..805L}. 
However, the necessary and sufficient conditions for the onset of the RWI have been unknown until recent years. 
In \citet[][hereafter paper I]{2016ApJ...823...84O}, we performed the linear stability analyses of the RWI within the framework of two-dimensional (2D), barotropic and purely hydrodynamic disks. 
We have revealed the parameter sets where the disks are unstable against the RWI and derived the critical condition for the RWI in a semi-analytic form. 
The RWI has been also studied with numerical simulations \citep{2001ApJ...551..874L, 2006A&A...446L..13V, 2008A&A...491L..41L, 2009A&A...497..869L, 2010A&A...516A..31M, 2012A&A...542A...9M, 2013A&A...559A..30R}. 
However, our knowledge on the evolution and the final outcomes of the RWI is still limited, partially due to lack of systematic parameter survey in 2D cylindrical coordinates. 
In this paper, we perform numerical simulations of the RWI in 2D, barotropic and purely hydrodynamic disks. 
We explore a wide parameter space of initial surface density profiles and the disk temperature. 
We investigate the RWI evolution, and the properties and evolution of the vortices formed by the RWI. 
%check 2

This paper is organized as follows. 
We describe our disk model and numerical setup in Section 2. 
We present results and discussions on the RWI evolution in Section 3 and on the vortices formed by the RWI in Section 4. 
Section 5 is for the summary. 
%check 2

\section{Disk Model and Numerical Method} \label{sec:method}
%Basic Equations
%Initial Conditions
\subsection{Disk Models and Initial Condition}

We consider barotropic and purely hydrodynamic disks orbiting a central star of mass $M$ in global 2D cylindrical coordinates, which are the same as the model used in our linear stability analyses presented in paper I. 
We assume that the disks are geometrically thin and neglect the effects of magnetic fields, viscosity, and self-gravity. 
We employ these rather simple assumptions in order to compare the numerical simulations with the results of paper I in detail and to perform a systematically controlled parameter survey. 
Previous works showed that viscosity, self-gravity, an indirect term of gravity force, vertical stratification and baroclinicity have some effects on the RWI or the vortices formed by the RWI \citep{2012ApJ...754...21L, 2012MNRAS.426.3211L, 2013ApJ...765...84L, 2013MNRAS.429..529L, 2014MNRAS.437..575L, 2016MNRAS.458.3918Z, 2016MNRAS.457.1944M, 2017ApJ...835..118M}. 
The numerical calculations of the RWI with dust particles, planets, and magnetic fields have been also performed \citep{2005ApJ...624.1003L, 2006ApJ...649..415I, 2011MNRAS.415.1426L, 2011MNRAS.415.1445L, 2012ApJ...756...62L, 2013MNRAS.433.2626R, 2014ApJ...788L..41F, 2014ApJ...795L..39F, 2017MNRAS.466.3533H}. 
However, we can capture the essential physics of the RWI even within the 2D, barotropic, and purely hydrodynamic framework. 
%check 2

Our numerical simulations employ a non-rotating frame centered on a star and a 2D cylindrical coordinate with ($r$, $\varphi$). 
The gravitational potential of the central star is given by $\Phi(r)\ =\ -GM/r$, where $G$ is the gravitational constant. 
We denote the surface density by $\Sigma$ and the (vertically integrated) pressure by $P$. 
We assume that the disk is barotropic, i.e., $P\ =\ P(\Sigma)\ \propto\ \Sigma^\Gamma$, where $\Gamma$ is the effective adiabatic index of the gas. 
In our simulations, we consider only $\Gamma \ =\ 5/3$. 
The continuity equation is 
\begin{equation}
\henb{\Sigma}{t} + \frac{1}{r}\henb{}{r} (r \Sigma v_r)+\frac{1}{r}\henb{}{\varphi}(\Sigma v_\varphi)=0, \label{eq:cont}
\end{equation}
where $t$ is the time, $\BF{v}(r,\ \varphi,\ t) \ \equiv \ v_r (r,\ \varphi,\ t)\ {\bf \hat{r}}\ +\ v_\varphi (r,\ \varphi, \ t)\ \hat{ \bf \varphi}$ is the velocity field, ${\bf \hat{r}}$ is the unit vector in the $r$ direction, and ${\bf \hat{\varphi}}$ is the unit vector in the $\varphi$ direction. 
The equations of motion are 
\begin{eqnarray}
\henb{v_r}{t} + v_r\henb{v_r}{r}+\frac{v_\varphi}{r}\henb{v_r}{\varphi}-\frac{v_\varphi^2}{r}&=&-\frac{GM}{r^2}-\henb{\Pi}{r}, \label{eq:mom1} \\
\henb{v_\varphi}{t} + v_r\henb{v_\varphi}{r}+\frac{v_\varphi}{r}\henb{v_\varphi}{\varphi}+\frac{v_r v_\varphi}{r}&=&-\frac{1}{r}\henb{\Pi}{\varphi}, \label{eq:mom2}
\end{eqnarray}
where $\Pi$ is the pressure function. 
For the barotropic flow with $\Gamma \ \neq \ 1$, $\Pi$ is written as 
\begin{equation}
\Pi \equiv \frac{\Gamma}{\Gamma-1}\frac{P}{\Sigma}.
\end{equation}
From equations (\ref{eq:cont})--(\ref{eq:mom2}), the equation of the vortensity conservation is obtained as 
\begin{equation}
\henb{q}{t}+v_r \henb{q}{r} +\frac{v_\varphi}{r}\henb{q}{\varphi}=0, 
\end{equation}
where $q(r,\ \varphi,\ t)\ \equiv \ (\mathrm{rot}\, {\bf v})_z/\Sigma$ is the vortensity. 
%check 2

We perform numerical calculations with various initial conditions and investigate the RWI and vortices formed by the RWI. 
We adopt stationary ($\partial /\partial t \ =\ 0$), axisymmetric ($\partial/ \partial \varphi \ =\ 0$), and circular ($v_r \ =\ 0$) flow as the initial conditions, which are denoted by subscripts `0', e.g., $\Sigma_0(r)$, $P_0(r)$ and $\BF{v}_0 (r)\ =\ v_{\varphi 0}(r)\ \hat{\BF{\varphi}}$. 
The initial surface density $\Sigma_0(r)$ is given by a Gaussian bump on a uniform profile: 
\begin{equation}
\frac{\Sigma_0}{\Sigma_\mathrm{n}} =1+\mathcal{A}_0 \exp \left[-\frac{1}{2}\left(\frac{r-r_\mathrm{n}}{\Delta w_0} \right)^2 \right], \label{eq:fit}
\end{equation}
where 
$\Sigma_\mathrm{n}$ is the surface density of the uniform profile and $r_\mathrm{n}$ is the representative radius of the initial bump. 
This initial profile is the same as the ``GB'' type profile in paper I. 
There are two parameters to characterize the initial bump profile $\Sigma_0$: the contrast $\mathcal{A}_0$ and the radial half-width $\Delta w_0$. 
%check 2

Since we consider the barotropic flow, $P_0(r)$ follows 
\begin{equation}
P_0(r)=S_\mathrm{0} \Sigma_0^\Gamma, \label{eq:eos}
\end{equation}
where $S_\mathrm{0}$ is the entropy and constant. 
We define a dimensionless parameter $h$ by 
\begin{equation}
h\equiv \frac{\sqrt{\Gamma S_0 \Sigma_\mathrm{n}^{(\Gamma-1)}}}{r_\mathrm{n} \Omega_\mathrm{n}},
\end{equation}
where $\Omega_\mathrm{K}(r) \ \equiv \ \sqrt{GM/r^3}$ is the Kepler angular velocity and $\Omega_\mathrm{n} \ \equiv \ \Omega_\mathrm{K}(r_\mathrm{n})$. 
In this case, the initial entropy is written as 
\begin{equation}
S_0=\frac{h^2}{\Gamma}\frac{(r_\mathrm{n} \Omega_\mathrm{n})^2}{\Sigma_\mathrm{n}^{(\Gamma-1)}}=\mathrm{constant}. 
\end{equation}
It is noted that $h$ can be regarded as the dimensionless disk scale-height, or, equivalently, the dimensionless sound speed. 
The value of $h$ also represents the disk temperature. 
From equation (\ref{eq:mom1}), the initial velocity field in the azimuthal direction $v_{\varphi 0} (r)$ is obtained as
\begin{equation}
v_{\varphi 0}(r)=\sqrt{v_\mathrm{K}^2+r\diff{\Pi_0}{r}}, 
\end{equation}
where $\Pi_0(r)\ \equiv \ \Gamma S_0 \Sigma_0^{\Gamma-1}/(\Gamma-1)$ is the initial pressure function. 
%check 2

The initial conditions are characterized by three parameters: $h$, $\Delta w_0$, $\mathcal{A}_0$. 
First, we fix $h$ and $\Delta w_0$ and vary $\mathcal{A}_0$. 
The larger $\mathcal{A}_0$ is, the more unstable against the RWI the system is.
For an unstable configuration against the RWI, the largest linear growth rate of the RWI, $\gamma_\ast (\mathcal{A}_0:\ h,\ \Delta w_0)$, monotonically increases with $\mathcal{A}_0$ (see paper I). 
If, however, $\mathcal{A}_0$ exceeds a certain value, $\mathcal{A}_{0, \mathrm{max}}(h,\ \Delta w_0)$, the system violates the Rayleigh's condition and is prone to the rotational instability (see Appendix B.1). 
Since the linear growth rate of the rotational instability is typically larger than that of the RWI, we expect that the system which is unstable against the rotational instability immediately transfers to the marginally stable configuration of the rotational instability ($\mathcal{A}_0\ =\ \mathcal{A}_{0, \mathrm{max}}(h,\ \Delta w_0)$). 
We, therefore, consider the cases where the system does not violate the Rayleigh's condition. 
In other words, we consider the cases with $\mathcal{A}_0\ < \ \mathcal{A}_{0, \mathrm{max}}(h,\ \Delta w_0)$ as the initial conditions. 
The maximum of the largest linear growth rate of the RWI is limited below the value of that with $\mathcal{A}_0\ = \ \mathcal{A}_{0, \mathrm{max}}(h,\ \Delta w_0)$, which we denote by $\gamma_{\ast, \mathrm{max}}(\mathcal{A}_{0, \mathrm{max}}:\ h,\ \Delta w_0)$. 
%check 2

\begin{table*}
\caption{The parameter sets of the models. \label{tbl:ini}}
\begin{center}
\renewcommand{\arraystretch}{1.0}
\hspace*{-7em}
\begin{tabular}[t]{l @{\hspace{-3em}} l}
\begin{tabular}[t]{lccccc}
\tableline \tableline
Name & $h$ & $\Delta w_0/r_\mathrm{n}$ &$\gamma_\ast/\Omega_\mathrm{n}$ & $\mathcal{A}_0$&$m_\ast$  \\
\hline 
\hline 
h10w1g1&0.1&2.00E$-$2&0.227&4.04E$-$2&9\\
h10w1g2&0.1&2.00E$-$2&0.200&3.40E$-$2&8\\
h10w1g3&0.1&2.00E$-$2&0.150&2.41E$-$2&7\\
h10w1g4&0.1&2.00E$-$2&0.100&1.57E$-$2&6\\
h10w1g5&0.1&2.00E$-$2&0.050&8.70E$-$3&4\\
h10w2g1&0.1&3.56E$-$2&0.242&1.31E$-$1&6\\
h10w2g2&0.1&3.56E$-$2&0.201&1.06E$-$1&5\\
h10w2g3&0.1&3.56E$-$2&0.150&7.81E$-$2&5\\
h10w2g4&0.1&3.56E$-$2&0.100&5.46E$-$2&4\\
h10w2g5&0.1&3.56E$-$2&0.050&3.42E$-$2&3\\
h10w3g1&0.1&6.32E$-$2&0.246&4.39E$-$1&4\\
h10w3g2&0.1&6.32E$-$2&0.200&3.58E$-$1&4\\
h10w3g3&0.1&6.32E$-$2&0.150&2.77E$-$1&3\\
h10w3g4&0.1&6.32E$-$2&0.100&2.05E$-$1&3\\
h10w3g5&0.1&6.32E$-$2&0.050&1.42E$-$1&2\\
h10w4g1&0.1&1.12E$-$1&0.227&1.57E$+$0&3\\
h10w4g2&0.1&1.12E$-$1&0.200&1.38E$+$0&2\\
h10w4g3&0.1&1.12E$-$1&0.150&1.06E$+$0&2\\
h10w4g4&0.1&1.12E$-$1&0.100&8.02E$-$1&2\\
h10w4g5&0.1&1.12E$-$1&0.050&6.00E$+$1&2\\
h10w5g1&0.1&2.00E$-$1&0.191&5.66E$+$0&2\\
h10w5g3&0.1&2.00E$-$1&0.150&4.69E$+$0&2\\
h10w5g4&0.1&2.00E$-$1&0.100&3.39E$+$0&1\\
h10w5g5&0.1&2.00E$-$1&0.050&2.33E$+$0&1\\
\tableline \tableline
\end{tabular}
&
\begin{tabular}[t]{lccccc}
\tableline \tableline
Name & $h$ & $\Delta w_0/r_\mathrm{n}$ &$\gamma_\ast/\Omega_\mathrm{n}$ & $\mathcal{A}_0$&$m_\ast$  \\
\hline 
\hline 
h20w1g1&0.2&2.00E$-$2&0.209&1.00E$-$2&8\\
h20w1g4&0.2&2.00E$-$2&0.100&3.97E$-$3&4\\
h20w2g1&0.2&3.56E$-$2&0.225&3.17E$-$2&5\\
h20w2g4&0.2&3.56E$-$2&0.100&1.23E$-$2&3\\
h20w3g1&0.2&6.32E$-$2&0.237&1.00E$-$1&3\\
h20w3g4&0.2&6.32E$-$2&0.100&4.15E$-$2&3\\
h20w4g1&0.2&1.12E$-$1&0.237&3.17E$-$1&2\\
h20w4g4&0.2&1.12E$-$1&0.100&1.53E$-$1&2\\
h20w5g1&0.2&2.00E$-$1&0.192&9.79E$-$1&2\\
h20w5g4&0.2&2.00E$-$1&0.100&5.38E$-$1&1\\ \hline
h15w1g1&0.15&2.00E$-$2&0.216&1.78E$-$2&8\\
h15w1g4&0.15&2.00E$-$2&0.100&6.89E$-$3&5\\
h15w2g1&0.15&3.56E$-$2&0.233&5.69E$-$2&5\\
h15w2g4&0.15&3.56E$-$2&0.100&3.56E$-$2&3\\
h15w3g1&0.15&6.32E$-$2&0.240&1.83E$-$1&4\\
h15w3g4&0.15&6.32E$-$2&0.100&8.13E$-$1&2\\
h15w4g1&0.15&1.12E$-$1&0.233&6.00E$-$1&2\\
h15w4g4&0.15&1.12E$-$1&0.100&2.96E$-$1&2\\
h15w5g1&0.15&2.00E$-$1&0.193&1.96E$+$0&2\\
h15w5g4&0.15&2.00E$-$1&0.100&1.13E$+$0&1\\ \hline
h05w1g1&0.05&2.00E$-$2&0.244&1.68E$-$1&11\\
h05w1g4&0.05&2.00E$-$2&0100&7.14E$-$2&7\\
h05w2g1&0.05&3.56E$-$2&0.248&5.86E$-$1&7\\
h05w2g4&0.05&3.56E$-$2&0.100&2.77E$-$1&5\\
h05w3g1&0.05&6.32E$-$2&0.240&2.30E$+$0&4\\
h05w3g4&0.05&6.32E$-$2&0.100&1.18E$+$0&4\\
h05w4g1&0.05&1.12E$-$1&0.225&1.01E$+$1&3\\
h05w4g4&0.05&1.12E$-$1&0.100&5.28E$+$0&2\\
h05w5g1&0.05&2.00E$-$1&0.188&4.12E$+$1&2\\
h05w5g4&0.05&2.00E$-$1&0.100&2.52E$+$1&1\\ \hline
\tableline \tableline
\end{tabular}
\\
\\
\multicolumn{2}{p{\textwidth}}{\hspace{3em} NOTE. Name: the name of the model. $h$: the dimensionless disk aspect ratio. $\Delta w_0/r_\mathrm{n}$: the radial}\\
\multicolumn{2}{p{\textwidth}}{\hspace{3em} half-width of the initial bump normalized by $r_\mathrm{n}$. $\gamma_\ast/\Omega_\mathrm{n}$: the largest linear growth rate against the}\\
\multicolumn{2}{p{\textwidth}}{\hspace{3em} RWI normalized by $\Omega_\mathrm{n}$. $\mathcal{A}_0$: the radial surface density contrast of the initial bump. $m_\ast$: the most}\\
\multicolumn{2}{p{\textwidth}}{\hspace{3em} unstable azimuthal mode.}
\end{tabular}
\end{center}
\end{table*}

When the three parameters $(h,\ \Delta w_0,\ \gamma_\ast)$ are given, $\mathcal{A}_0$ is uniquely determined. 
We vary $h$ and $\Delta w_0$ in the ranges of $h\ =\ [0.05,\ 0.1,\ 0.15,\ 0.2]$ and $\Delta w_0/r_\mathrm{n}\ =\ [0.02,\ 0.0356,\ 0.0632,\ 0.112,\ 0.2]$. 
We also vary $\gamma_\ast$ in the ranges of $\gamma_\ast / \Omega_\mathrm{n} \ =\ [\gamma_{\ast, \mathrm{max}}/ \Omega_\mathrm{n},\ 0.2,\ 0.15,\ 0.1,\ 0.05]$ for $h\ =\ 0.1$ and $\gamma_\ast / \Omega_\mathrm{n} \ =\ [\gamma_{\ast, \mathrm{max}} / \Omega_\mathrm{n},\ 0.1]$ for $h\ \neq\ 0.1$. 
We run 54 models in total whose the parameter sets are shown in Table \ref{tbl:ini}. 
Note that we do not have the ``h10w5g2'' model because $\gamma_{\ast, \mathrm{max}} / \Omega_\mathrm{n}$ is smaller than $0.2$ for $h\ =\ 0.1$ and $\Delta w_0\ =\ 0.2r_\mathrm{n}$. 
We show the most unstable azimuthal mode $m_\ast$ as well as the largest linear growth rate $\gamma_\ast$ in Table \ref{tbl:ini}. 
In addition, we calculate the linear growth rate of the RWI for each azimuthal mode $m$, $\gamma_m$, in the same manner as described in paper I. 
The setup of the linear stability analyses and the linear growth rates for $1\ \leq \ m\ \leq \ 10$ are shown in Appendix C.1. 
In this paper, we regard the ``h10w3g1'' model ($h\ =\ 0.1,\ \Delta w_0\ =\ 0.0632r_\mathrm{n}$,\ $\gamma_\ast / \Omega_\mathrm{n} \ =\ 0.246$, $ \mathcal{A}_0\ =\ 0.439$, and $m_\ast\ =\ 4$) as a fiducial case. 
When we investigate the overall properties of the RWI and vortices formed by the RWI, we always refer to the outcome of the ``h10w3g1'' model. 
%check 2

%Numerical Method
\subsection{Numerical Method}

We use the Athena++ code \citep{STW}, with the HLLC approximate Riemann Solver, the second-order piece-wise linear reconstruction, and the second-order van-Leer time integrator. 
We assume barotropic flows for simplicity and therefore we overwrite the pressure after every time step to satisfy equation (\ref{eq:eos}). 
The computational domain extends $r_\mathrm{in}\ <\ r\ <\ r_\mathrm{out}$, where we set $r_\mathrm{in}\ =\ 0.3~r_\mathrm{n}$ and $r_\mathrm{out}\ =\ 2.5~r_\mathrm{n}$, in the radial direction and covers full $2\pi$ in the azimuthal direction in the 2D cylindrical coordinates. 
We choose the radial extension of the numerical domain so that all the effective Lindblad resonances from the co-rotation point with the vortex center reside within the computational domain unless vortices become too close to the boundaries. 
%check 2

We set the mesh structure so that the size of a cell is at least smaller than $0.04h$ in the vortex-forming region ($r\ \sim \ r_\mathrm{n}$). 
For $h\ =\ 0.1,\ 0.15,$ and $0.2$, the mesh has 576 cells in the radial direction and 1596 grids in the azimuthal direction. 
For $h\ =\ 0.05$, the mesh has 1296 cells in the radial direction and 3744 cells in the azimuthal direction. 
While the azimuthal spacing of cells is uniform, we make the radial spacing logarithmically constant and keep an aspect ratio of cells about unity. 
From the resolution study, the calculations with the mesh structure are high-resolution enough to discuss the results of this paper (see Appendix D). 
%check 2

We adopt the non-reflective boundary conditions \citep{1996MNRAS.282.1107G} in the radial direction and the periodic boundary conditions in the azimuthal direction. 
The non-reflective boundary conditions are designed to be non-reflective only for one-dimensional simple waves. 
Even for 2D nonlinear simulations, however, we have observed the strong reduction of the wave reflection at the radial boundaries. 
This non-reflective boundary conditions are also used in previous works \citep[e.g.,][]{2010ApJ...725..146P}. 
Note that this non-reflective boundary conditions cannot vanish the wave reflection perfectly. 
However, the inner boundary does not have significant effects on the vortices formed by the RWI (see Appendix D). 
%check 2

We have further modified the original Athena++ code by introducing fast Fourier Transform (FFT) filters. 
We perform the Fourier transform of $\Sigma, \ \mathcal{M}_r$, and $\mathcal{M}_\varphi$ in the azimuthal direction at every radius and every time step. 
We denote the Fourier components for the azimuthal mode $m$ by $\mathcal{F}(\Sigma)_{m}, \ \mathcal{F}(\mathcal{M}_{r})_m,$ and $\mathcal{F}(\mathcal{M}_{\varphi})_m$. 
We have developed two kinds of FFT filters, namely, ``axisymmetric filter" and ``single-mode filter". 
For the numerical relaxation of the initial conditions before the main calculations, we use the axisymmetric filter, where all the non-axisymmetric ($m\ \neq \ 0$) modes are filtered out. 
We use the single-mode filter, where all the non-axisymmetric components except for $m\ =\ k$ are filtered out when we investigate the linear and the weakly non-linear regimes and the saturation of a specific azimuthal mode $k$ (see Section 3.1 and Section 3.2). 
After these filters modify the Fourier components, we recalculate $\Sigma , \ \mathcal{M}_r$, and $\mathcal{M}_\varphi$ by the inverse Fourier transform and update the quantities in the calculations. 
%check 2

Before starting the main calculation of each run, we evolve the disk numerically using the axisymmetric filter for 10 orbits at $r\ =\ r_\mathrm{n}$ in order to relax the initial profile to a numerical equilibrium. 
We impose a small initial perturbation on the radial momentum, $\mathcal{M}_{r}$, to trigger the RWI and start the main calculation. 
The Fourier component of the radial momentum for an azimuthal mode $m$ is defined by $\mathcal{F}(\mathcal{M}_{r})_m$. 
We perform two types of numerical calculations. 
The first is a single-mode calculation in which we focus on one specific azimuthal mode $k$. 
In the single-mode calculations, the initial perturbations satisfy $|\mathcal{F}(\mathcal{M}_{r})_m|\ =\ 10^{-6}\, |\mathcal{F}(\mathcal{M}_{\varphi})_{m=0}| \exp[\{(r/r_\mathrm{n}-1)/0.2\}^2/2]$ for $m\ = \ k$ and $|\mathcal{F}(\mathcal{M}_{r})_{m=0}|\ =\ 0$ otherwise, where $\mathcal{M}_{\varphi}$ is the azimuthal momentum and $\mathcal{F}(\mathcal{M}_{\varphi})_{m=0}$ is the axisymmetric Fourier component of the azimuthal momentum. 
At that point, the $k$ mode is {\it not} restricted to the most unstable azimuthal mode. 
We also make the single-mode filter of the $k$ mode work to filter out the other non-axisymmetric components ($m\ \neq \ k$) during calculations. 
We use the results of the single-mode calculations for the purpose of investigating the initial evolution and saturation of the RWI. 
The other is a white-noise calculation. 
In the white-noise calculations, the initial perturbations satisfy $|\mathcal{F}(\mathcal{M}_{r})_m|\ =\ 10^{-6}\ |\mathcal{F}(\mathcal{M}_{\varphi})_{m=0}| \exp[\{(r/r_\mathrm{n}-1)/0.2\}^2/2]$ for $1\ \leq \ m \ \leq \ 128$. 
Note that we set a maximum azimuthal mode of the white-noise perturbation to $m\ =\ 128$ in order to avoid the effects from the numerical resolution. 
The phase of $\mathcal{F}(\mathcal{M}_{r})_m$ is randomly varied for each $m$. 
In the white-noise calculations, we do not use the single-mode filter. 
If not stated otherwise, we refer to the white-noise calculations. 
%check 2

\section{Evolution of the RWI} \label{sec:overview}

\begin{figure*}
\plotone{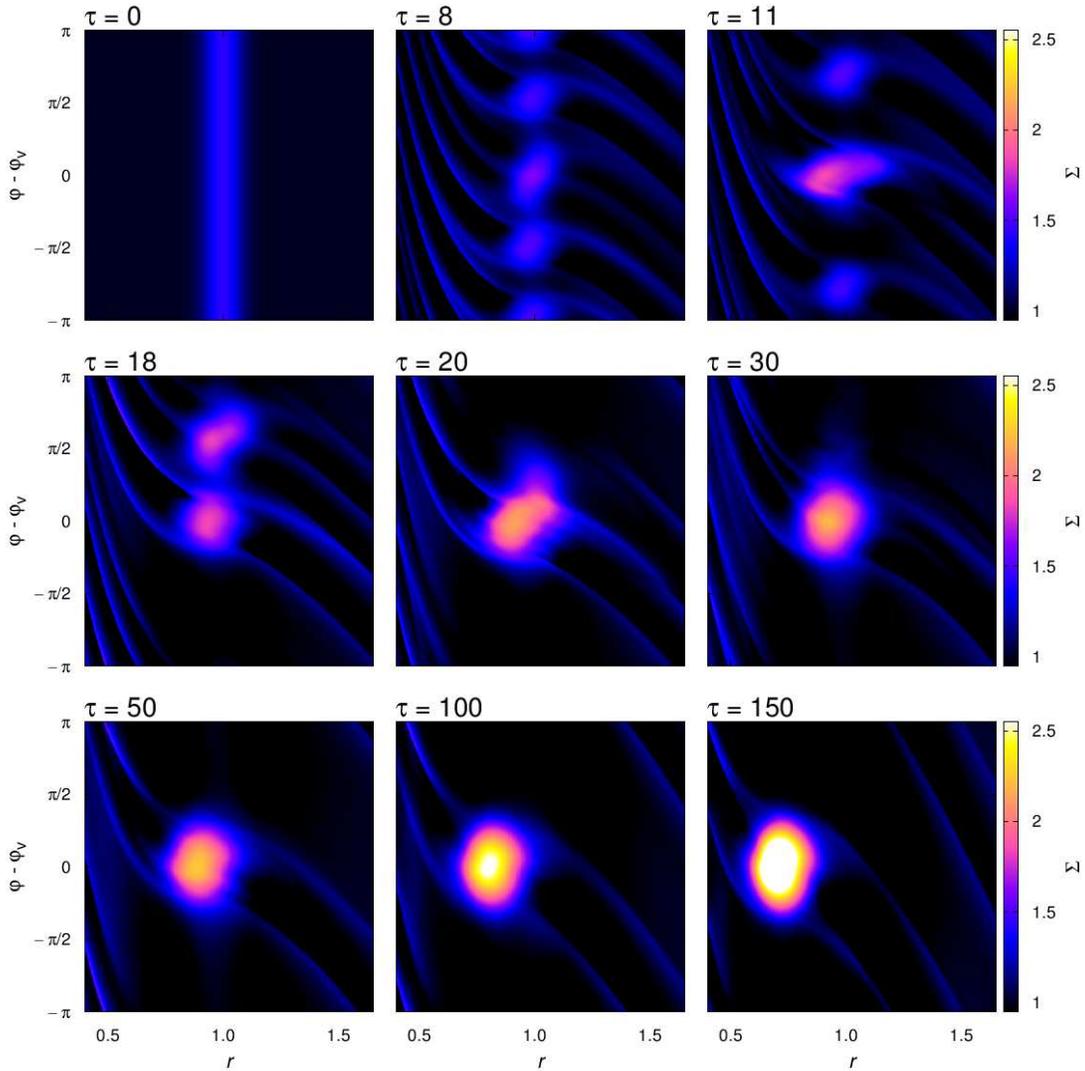}
\vspace{-0.15cm}
\caption{The snapshots of the surface density at $\tau \ =\ 0,\ 8,\ 11,\ 18,\ 20,\ 30,\ 50,\ 100$ and $150$ in the white-noise calculation of the ''h10w3g1'' model.}
\label{fig:2d}
\end{figure*}

First of all, we give an overview of the RWI evolution. 
Figure \ref{fig:2d} shows 2D snapshots of the surface density at $\tau \ =\ 0,\ 8,\ 11,\ 18,\ 20,\ 30, 50,\ 100$ and 150 in the fiducial calculation, where $\tau \ \equiv \ t\Omega_\mathrm{n}/2\pi$ is the time measured in the unit of the orbital period at $r\ =\ r_\mathrm{n}$. 
%check 2

%check2

After the onset of the RWI, the perturbation shows linear and weakly non-linear evolution. 
The saturation occurs when the perturbation becomes comparable to the initial axisymmetric bump. 
And then the system enters the fully non-linear regime at $\tau \ \sim \ 8.3$. 
At that time, four vortices are formed by fragmentation of the initial axisymmetric bump. 
The number of the vortices formed initially is in accordance with the most unstable azimuthal mode of the RWI, $m_\ast$. 
The vortices coalesce one after another ($4 \ \rightarrow \ 3$ at $\tau \ \sim \ 11$; $3\ \rightarrow \ 2$ at $\tau \ \sim \ 18$; $2 \ \rightarrow \ 1$ at $\tau \ \sim \ 20$). 
In the end, one quasi-stationary vortex remains after the final merger. 
%check 2

In this section, we consider each stage of the RWI evolution individually: the linear and weakly non-linear regimes in Section 3.1, the saturation in Section 3.2, and the vortex merger in Section 3.3. 
%check 2

\begin{figure*}
\plotone{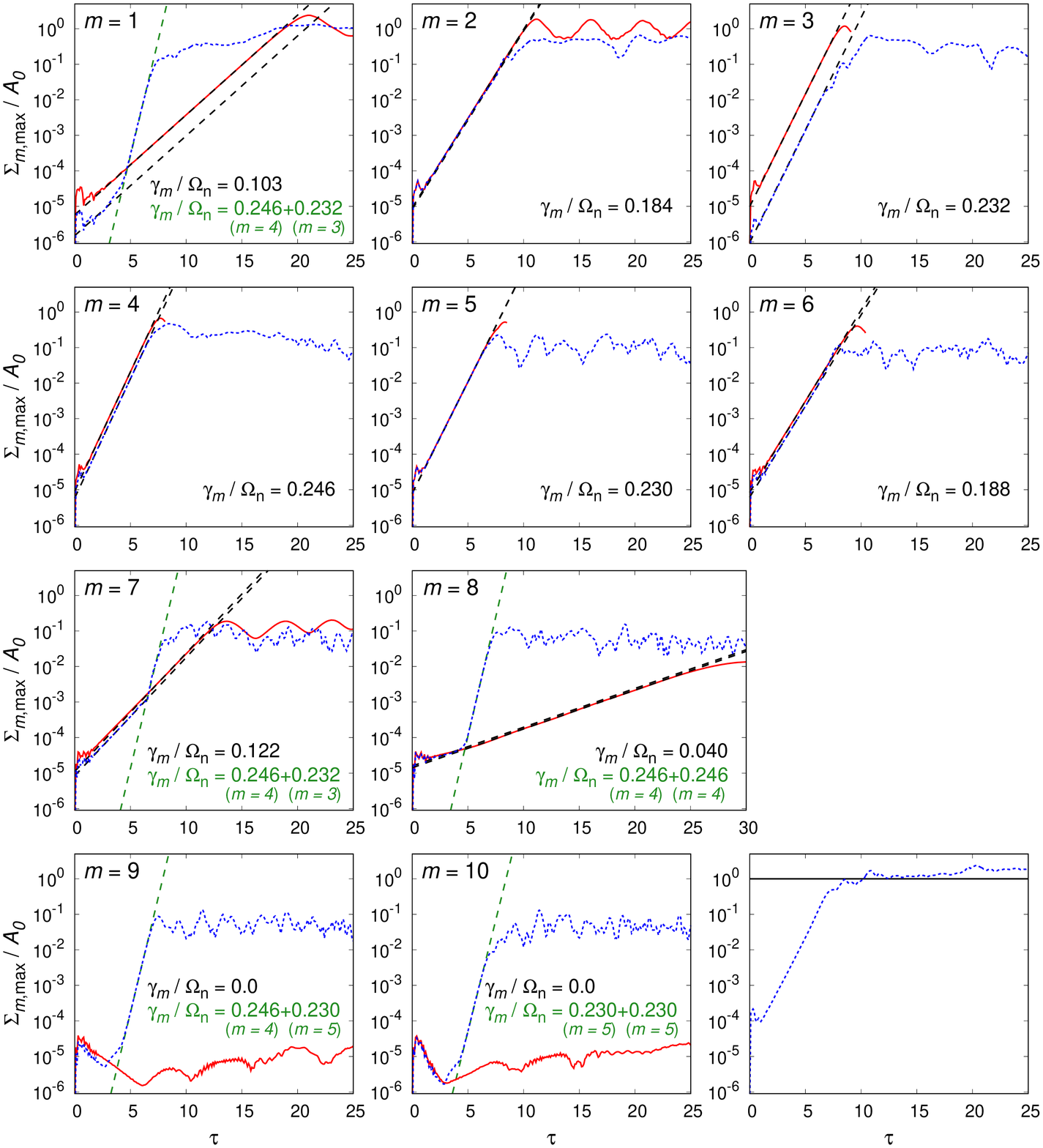}
\vspace{-1.3cm}
\caption{The time evolution of $\Sigma_{m,\mathrm{max}}$ in the single-mode calculations (the red lines) and the white-noise calculation (the blue dotted lines) of the ``h10w3g1'' model for each azimuthal mode $m$ ($1\ \leq \ m\ \leq \ 10$). The black dashed lines show the linear growth and the green dashed lines show the growth estimated from the mode-mode coupling based on the linear stability analyses. The last Panel shows the time evolution of the maximum of $\sum_{m\geq 1} \Sigma_m$ in the white-noise calculation (the blue dotted line).}
\label{fig:single}
\end{figure*}
%check2

\begin{figure*}
\epsscale{0.92}
\plotone{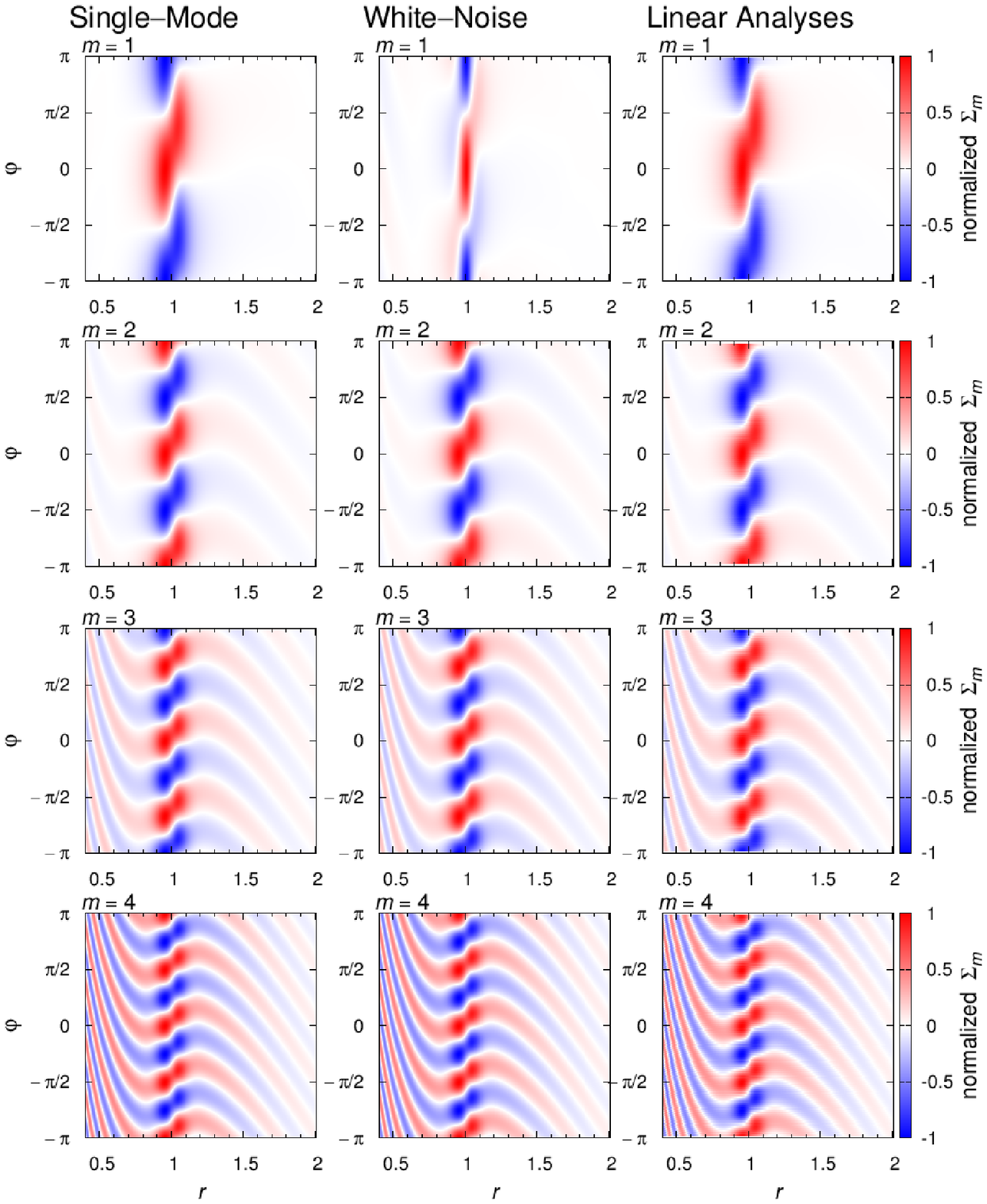}
\caption{The distribution of the normalized surface density perturbations in the single-mode calculations (the left column) and the white-noise calculation (the middle column) of the ``h10w3g1'' model at $\tau~=~5.6$ for each azimuthal mode $m$ ($1\ \leq \ m\ \leq \ 4$, from the top row to the bottom). The right panels show the distribution of the surface density perturbations derived from the linear stability analyses for each $m$.}
\label{fig:linear}
\end{figure*}
%check2

\subsection{Linear and Weakly Non-linear Regime of the RWI} 

Here, we pay our attention to the linear and weakly non-linear regimes of the RWI. 
We compare the results of the numerical calculations with those of the linear stability analyses in order to confirm the validity of our numerical calculations. 
We also take our step into the weakly non-linear regime and study how applicable the linear stability analyses are to understand the RWI evolution. 
%check 2

For the sake of the comparison with the linear stability analyses, we separate the surface density in the numerical calculations into axisymmetric components and non-axisymmetric components. 
The axisymmetric component corresponds to the azimuthally averaged surface density $\average{\Sigma}(r)$. 
We define the non-axisymmetric component of the $m$ mode by $\Sigma_m(r, \ \varphi)\ \equiv \ \mathrm{Real}\left[ \mathcal{F}(\Sigma)_m\exp(im\varphi)\right]$. 
The single-mode calculation for the $m$ mode has only the $m$ mode component and the axisymmetric component. 
Therefore, the $m$ mode component is calculated by subtracting $\average{\Sigma}(r)$ from $\Sigma(r, \ \varphi)$. 
On the other hand, the white-noise calculation requires for using the single-mode filter in a post-process to obtain the non-axisymmetric components. 
%check 2

We perform single-mode calculations for the $m$ modes ($1\ \leq \ m\ \leq \ 10$) and a white-noise calculation of the ``h10w3g1'' model. 
Figure \ref{fig:single} shows the time evolution of $\Sigma_\mathrm{m, max}$ in the calculations, where $\Sigma_\mathrm{m, max}$ is the maximum of $\Sigma_m$ around $r\ =\ r_\mathrm{n}$. 
The linear growth rates against the RWI are independently derived from the linear stability analyses. 
All the numerical calculations show excellent agreement with the linear analyses in the linear regime. 
The white-noise calculation also shows the weakly non-linear growth of the mode with a small linear growth rate due to the coupling between the two modes with a large linear growth rate. 
For example, we can observe the mode coupling regime between the $m\ =\ 3$ mode and the $m\ =\ 4$ mode to produce the $m\ =\ 1\ (=\ 4-3)$ mode component in $4.4\ \leq \ \tau \ \leq \ 6.9$. 
We find that the linear stability analyses predict the weakly non-linear evolution precisely. 
%check 2

Figure \ref{fig:linear} compares the distribution of $\Sigma_m (r,\ \varphi)$ normalized by $\Sigma_{m, \mathrm{max}}$ in the single-mode calculations and the white-noise calculation at $\tau \ =\ 5.6$ and the surface density perturbation normalized by the maximum value derived in the linear stability analyses for $1\ \leq \ m \ \leq \ 4$. 
The azimuthal phase is shifted so that the point of $\Sigma_m \ = \ \Sigma_{m, \mathrm{max}}$ is at $\varphi \ =\ 0$. 
Note that we also denote the surface density perturbation of the linear stability analyses for the $m$ mode by $\Sigma_m (r,\ \varphi)$. 
Except for the $m\ = \ 1$ mode in the white-noise calculation, the profiles of $\Sigma_m$ in the numerical calculations match those in the linear stability analyses. 
The discrepancy for the $m\ =\ 1$ mode occurs because the coupling between the $m\ =\ 3$ mode and the $m\ =\ 4$ mode becomes significant and the $m\ =\ 1$ mode already enters the weakly non-linear regime at $\tau \ =\ 5.6$ in the white-noise calculation. 
%check 2

From Figures \ref{fig:single} and \ref{fig:linear}, our numerical calculations agree with the linear stability analyses in the linear and weakly non-linear regimes. 
Therefore, our numerical calculations and linear stability analyses are reliable in these regimes. 
%check 2

\subsection{Saturation Mechanism of the RWI} 

As shown in Figure \ref{fig:single}, the RWI saturation occurs when the amplitude of the non-axisymmetric components becomes comparable to $\mathcal{A}_0$ in both the single-mode calculations and the white-noise calculation. 
During the growth of the non-axisymmetric components, the axisymmetric components, or the $m\ =\ 0$ mode components, also evolve due to the couplings of the non-axisymmetric components. 
For example, a self-coupling of the $m\ =\ k$ mode can produce the $m\ =\ 0\ (=\ k-k)$ mode component. 
As another example, the couplings between three or more modes also can produce the $m\ =\ 0$ mode components. 
Here, we attempt to explain the saturation mechanism of the RWI investigating the time evolution of the axisymmetric components. 
%check 2

\begin{figure}
\epsscale{1.15}
\plotone{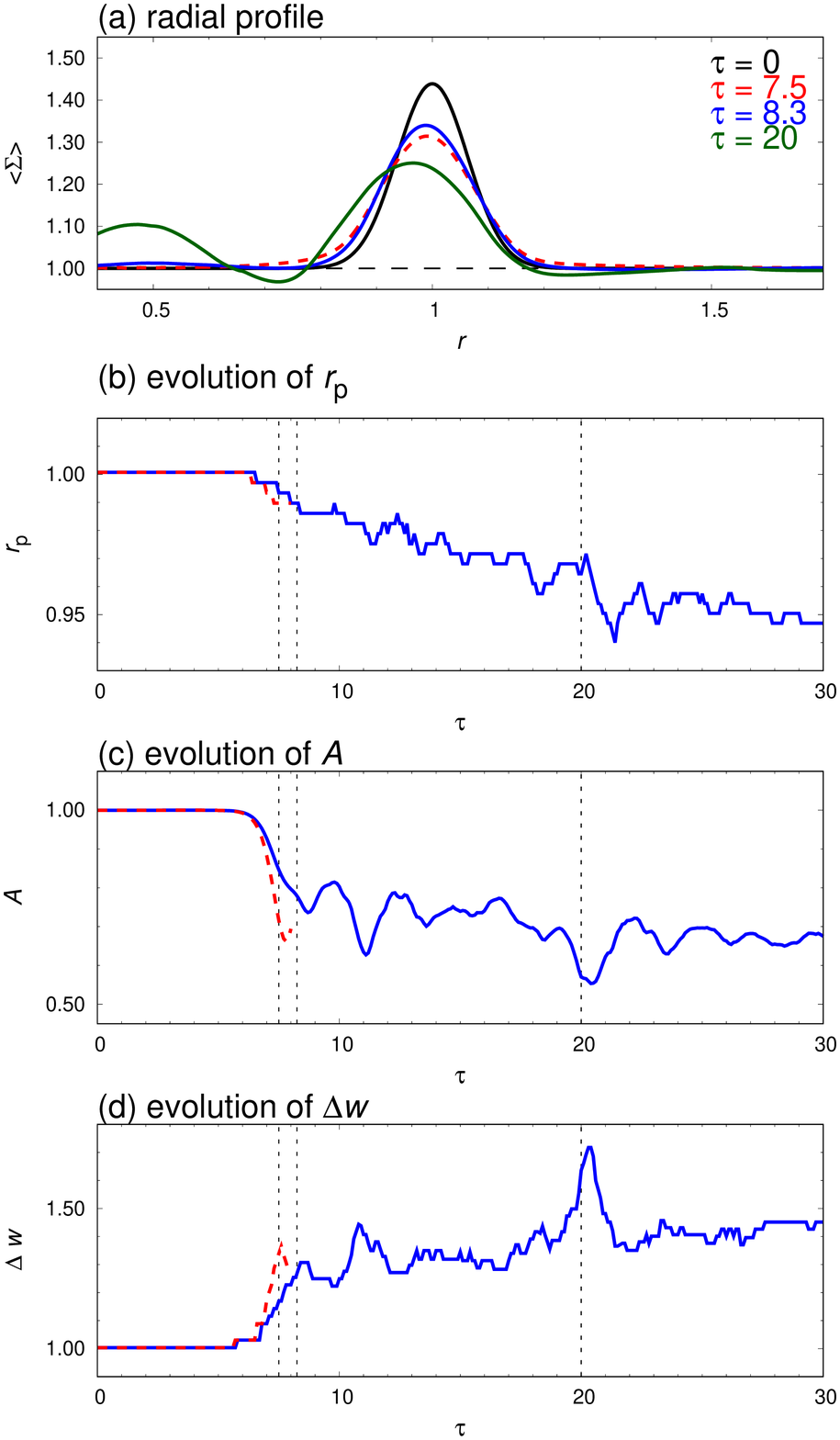}
\vspace{-0.35cm}
\caption{The time evolution of the azimuthally averaged profiles in the white-noise calculation (the solid lines) and the single-mode calculation for the $m\ =\ 4$ mode (the dashed lines) of the ``h10w3g1'' model. Panel (a) shows the initial surface density ($\tau \ =\ 0$) with the black solid line and the azimuthally averaged surface density at $\tau\ =\ 7.5$ (red), $8.3$ (red), and 20 (green). The time evolution of $r_p$, $\mathcal{A}$, and $\Delta w$ is shown in Panels (b), (c), and (d), respectively. \label{fig:res_ave}}
\end{figure}

We analyze the axisymmetric components in the single-mode calculation for the $m\ =\ 4$ mode and the white-noise calculation of the ``h10w3g1'' model. 
Since the radial profiles of the azimuthally averaged surface density $\average{\Sigma}(r)$ resemble a Gaussian bump during the RWI evolution as seen in Panel (a) of Figure \ref{fig:res_ave}, we measure the location of the peak $r_\mathrm{p}$, the contrast $\mathcal{A}$, and the half-width $\Delta w$ of the bump by fitting with $\average{\Sigma} / \Sigma_\mathrm{n}\ =\ \mathcal{A}\exp[-\{(r-r_\mathrm{p})/\Delta w\}^2/2]+1$. 
From Panels (b)--(d) of Figure \ref{fig:res_ave}, $r_\mathrm{p}$ and $\mathcal{A}$ start to decrease and $\Delta w$ starts to increase a few orbits before the saturation in both calculations. 
While the change of $r_\mathrm{p}$ is gradual, the changes of $\mathcal{A}$ and $\Delta w$ are rapid. 
These mean that the axisymmetric components approach the stable configurations against the RWI during the RWI evolution. 
%check 2

%check2

In order to investigate quantitatively the time evolution of the axisymmetric components, we use the semi-analytic condition for the onset of the RWI derived in paper I: 
\begin{equation}
\eta_m\equiv \int^{r_\mathrm{OR}}_{r_\mathrm{IR}}\sqrt{-D_{\mathrm{MS}, m}(r)}\mathrm{d}r \gtrsim \eta_\mathrm{c}, \label{eq:nsc}
\end{equation}
where $D_{\mathrm{MS}, m}$ is the effective potential of the $m$ mode if the system is assumed to be marginally stable against the RWI of the $m$ mode and $r_\mathrm{IR}$ and $r_\mathrm{OR}$ are the radii where $D_{\mathrm{MS}, m}$ vanishes. 
The threshold of the condition, $\eta_\mathrm{c}$, is roughly equal to $\pi/(2\sqrt{2})$ when the profile of $D_{\mathrm{MS}, m}$ for $r_\mathrm{IR}\ <\ r\ <\ r_\mathrm{OR}$ is approximated by a parabolic function. 
We show that the detailed expression for $D_{\mathrm{MS}, m}$ in Appendix B.2. 
Since $\eta_m$ depends on the azimuthal mode $m$ and the axisymmetric components, $\eta_m$ evolves with the axisymmetric components if $m$ is fixed. 
Calculating $\eta_m$ every one-tenth orbit, we study the time evolution of the stability of the axisymmetric components against the RWI for the $m$ mode. 
%check 2

\begin{figure}
\epsscale{1.15}
\plotone{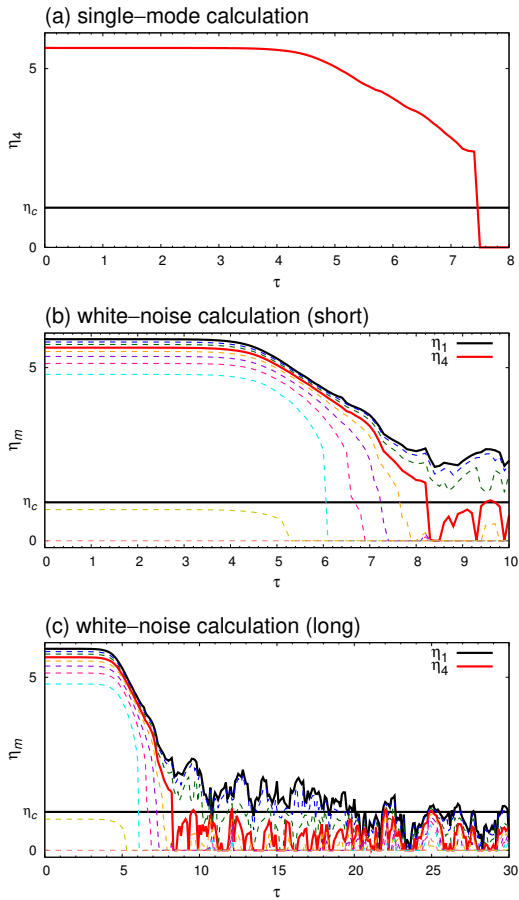}
\vspace{-1.0cm}
\caption{
The time evolution of the stability of the axisymmetric components against the RWI in the ''h10w3g1'' runs.  Panel (a) shows the time evolution of $\eta_4$ in the single-mode calculation for the $m\ =\ 4$ mode. The time evolution of $\eta_m$ for the $m\ =\ 1$ mode (the black solid lines), the $m\ =\ 4$ mode (the red solid lines), and other azimuthal modes $m\ \leq \ 10$ (the dashed lines) is shown for a short period ($\tau \ \leq \ 10$) in Panel (b) and for a long period ($\tau \ \leq \ 30$) in Panel (c). Here, $\eta_\mathrm{c} \ =\ \pi/(2\sqrt{2})$ is assumed. } 
\label{fig:eta1}
\end{figure}

First, we look at the time evolution of $\eta_4$ in the single-mode calculation for the $m\ =\ 4$ mode because the $m\ =\ 4$ mode is the most unstable azimuthal mode of the ``h10w3g1'' model. 
As shown in Panel (a) of Figure \ref{fig:eta1}, $\eta_4$ is initially larger than $\pi/(2\sqrt{2})$ so that the system is unstable against the RWI of the $m\ =\ 4$ mode. 
As the RWI evolves, $\eta_4$ decreases and becomes smaller than $\pi/(2\sqrt{2})$ at $\tau \ \approx \ 7.5$. 
This means that the axisymmetric component approaches the stable configuration during the RWI evolution and reaches the marginally stable configuration at the RWI saturation. 
The same thing occurs in the white-noise calculation, where the RWI is saturated at $\tau \ \approx \ 8.3$, even though the calculation contains all the non-axisymmetric components as well as the axisymmetric component as shown in Panel (b) of Figure \ref{fig:eta1}. 
Therefore, we consider that the RWI saturation occurs when axisymmetric components become marginally stable against the RWI for the most unstable azimuthal mode of the initial conditions. 
This indicates that the evolution of the axisymmetric components is mainly due to the self-coupling of the most unstable azimuthal mode. 
We also find that $\eta_{m>4}$ is smaller and $\eta_{m<4}$ is larger than $\pi/(2\sqrt{2})$ at the RWI saturation in the white-noise calculation. 
In other words, the axisymmetric components are stable for the higher modes but still unstable for the lower modes at the RWI saturation. 
%check 2

%check2

We also observe the similar evolution of $\eta_m$ in other calculations. 
However, the time when $\eta_{m=m_\ast}$ becomes smaller than $\pi/(2\sqrt{2})$ deviates from that of the RWI saturation in the models with a small linear growth rate. 
We consider that the time deviation is due to $\eta_\mathrm{c} \ \neq \ \pi/(2\sqrt{2})$ because the profile of $D_{\mathrm{MS}, m}$ for $r_\mathrm{IR}\ <\ r\ <\ r_\mathrm{OR}$ is not approximated very well by a parabolic function when the initial Gaussian bump is weak, i.e., the linear growth rate is small. 
%check 2

\citet{2013MNRAS.430.1988M} interpreted the saturation mechanism of the RWI in an analogy of the wave-particle interaction in plasma physics. 
Our explanation is based on the linear stability of the axisymmetric components and is complementary to that by \citet{2013MNRAS.430.1988M}. 
We expect that combining these explanations help us understand the physical mechanisms of the RWI evolution. 
%check 2

\subsection{Vortex Merger}

After the RWI saturation, multiple vortices formed as a result of the RWI coalesce one after another. 
In this section, we investigate the vortex merger regime. 

In all the runs, the regimes with more than two vortices continue at most for a few orbits. 
On the other hand, the lifetime of the two vortices regime shows some variations. 
By visual inspection of the surface density distribution, we identify the orbits when the vortex mergers occur. 
We define an orbit when the number of the vortices becomes two by $\tau_2$ and an orbit when the final vortex merger occurs by $\tau_1$. 
On that account, $\tau_2-\tau_1$ represents the duration of the two vortices regime. 
We show the values of $\tau_2$ and $\tau_1$ in Appendix C.2. 
Note that these orbits have errors of a few tenths due to the uncertainties of our visual inspection. 
For all the models with $m_\ast \ =\ 1$ and some models with $m_\ast \ =\ 2$, it is difficult to measure $\tau_2$ and $\tau_1$ so that we set $\tau_2$ to no data and $\tau_1$ to the orbit number at the RWI saturation. 
%check 2

\begin{figure}
\epsscale{1.1}
\plotone{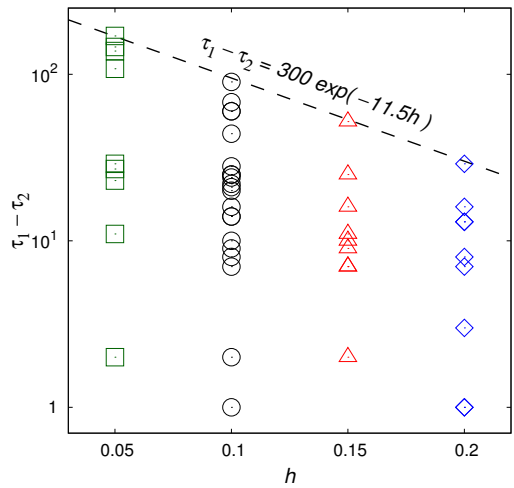}
\vspace{-0.55cm}
\caption{The duration of the two vortices regime. The values of $\tau_1-\tau_2$ are shown with the green squares ($h\ =\ 0.05$), the black circles ($h\ =\ 0.1$), the red triangles ($h\ =\ 0.15$), and the blue diamonds ($h\ =\ 0.2$), respectively. The black dashed line shows $\tau_1-\tau_2\ =\ 300\exp(-11.5h)$. \label{fig:tau}}
\end{figure}

The values of $(\tau_1-\tau_2)$ seem to be random. 
As shown in Figure \ref{fig:tau}, however, there is an upper limit in $(\tau_1-\tau_2)$; 
\begin{eqnarray}
\tau_1-\tau_2 &\lesssim& 300\exp(-11.5h), \nonumber \\
&\approx& 95 \exp[1-h/0.1], \label{eq:tau1}
\end{eqnarray}
within our parameter space. 
For $r_n\ =\ 100\ \mathrm{AU}$, one orbit corresponds to about $10^3\ \mathrm{yrs}$. 
From equation (\ref{eq:tau1}), the lifetime of the two vortices regime is up to about a few $\times \ 0.1\ \mathrm{Myr}$. 
The duration of the two vortices regime is one to two orders of magnitude shorter than the disk lifetime which is $1$--$10\ \mathrm{Myr}$ \citep[e.g.,][]{2001ApJ...553L.153H}. 
It is difficult to observe protoplanetary disks with multiple vortices formed by the same RWI event except at outer disks. 
%check 2

%check2

The vortex mergers strongly depend on the perturbations imposed on the initial conditions. 
In our calculations, the white-noise perturbations always have the same Fourier phase and the power spectrum because we use the same random seed and set the maximum azimuthal mode to $m\ =\ 128$. 
When the Fourier phase or power spectrum of the perturbations is different, the time when the vortex mergers occur varies. 
Even in those cases, however, $\tau_2 - \tau_1$ always satisfies equation (\ref{eq:tau1}). 
%check 2

We turn our attention to the stability of the axisymmetric components during the vortex merger regime. 
As discussed in Section 3.2, the axisymmetric components are still unstable at the RWI saturation for the lower modes than the most unstable azimuthal mode. 
After the saturation, $r_\mathrm{p}$ and $\mathcal{A}$ continue to decrease and $\Delta w$ increases. 
The rate of change of $r_\mathrm{p}$ is similar to that before saturation, but the rates of change of $\mathcal{A}$ and $\Delta w$ are slower. 
From Panel (c) of Figure \ref{fig:eta1}, the values of $\eta_m$ for $1\ \leq \ m\ \leq \ 3$ continue to decrease during the vortex mergers and finally become below the threshold. 
Particularly, $\eta_1$ reaches the threshold just after the final vortex merger at $\tau \ \approx \ 20$. 
Therefore, the axisymmetric components evolve toward the stable configurations during the vortex merger regime and become marginally stable against the RWI for the $m\ =\ 1$ mode at the final vortex merger. 
%check 2

\section{Quasi-Stationary Vortex Formed by the RWI}

We call the quasi-stationary vortex formed after the final vortex merger ``RWI vortex''. 
Hereafter, we focus on the structure and evolution of the RWI vortex. 
We provide the method to analyze the RWI vortex in Section 4.1 and show results in Section 4.2. 
Section 4.3 is for discussion. 
%check 2

\subsection{Analyses}
\subsubsection{Structure of the RWI vortex}
We show definitions of some physical quantities which characterize the vortex structure (vortex center, velocity gradient, vortex size, vortex aspect ratio, and turnover time). 
%check 2

\begin{figure*}
%\epsscale{1.0}
\plotone{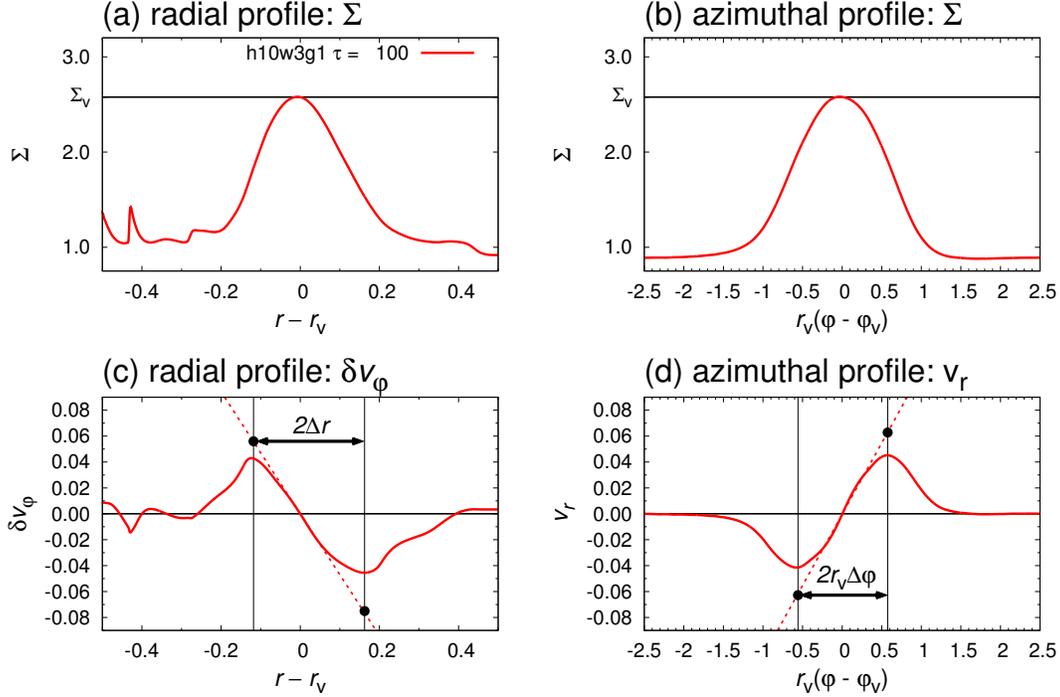}
\vspace{-0.35cm}
\caption{The structure of the RWI vortex in the ''h10w3g1'' run at $\tau \ =\ 100$. Panels (a)--(d) show the profiles of $\Sigma(r,\varphi_\mathrm{v})$, $\Sigma(r_\mathrm{v},\varphi)$, $\delta v_\varphi(r,\varphi_\mathrm{v})$, and $v_r(r_\mathrm{v},\varphi)$, respectively. The red dashed line in Panels (c) (or (d)) shows the radial (azimuthal) gradient of $\delta v_{\varphi}$ ($v_r$) at the vortex center. The black points show the values at the extrema expected from the velocity gradients. }
\label{fig:uzu1}
\end{figure*}

We define the center of the RWI vortex ($r_\mathrm{v}$, $\varphi_\mathrm{v}$) by 
\begin{eqnarray}
\delta v_{\varphi}(r_\mathrm{v}, \varphi_\mathrm{v}) &=& 0, \\
v_r(r_\mathrm{v}, \varphi_\mathrm{v}) &=& 0, 
\end{eqnarray}
in the vicinity of the surface density peak, where $\delta v_{\varphi}(r, \varphi)~\equiv~v_\varphi(r, \varphi)-v_\mathrm{K}(r)$. 
From Panels (a) and (b) of Figure \ref{fig:uzu1}, the surface density at the vortex center, $\Sigma_\mathrm{v} \ \equiv \Sigma(r_\mathrm{v}, \varphi_\mathrm{v})$, is almost the same as the peak value of the surface density. 
%check 2

%check2

Next, we turn our attention to the velocity field in the vortex and the vortex size. 
Panels (c) and (d) of Figure \ref{fig:uzu1} show the radial profile of $\delta v_\varphi$ and the azimuthal profile of $v_r$, respectively. 
In the vicinity of the vortex center, these velocity profiles are almost on the straight lines. 
We define the radial and azimuthal velocity gradients at the vortex center by 
\begin{eqnarray}
\delta v_{\varphi \mathrm{v},r}&=&\left[ \henb{\delta v_\varphi(r, \varphi_\mathrm{v})}{r}\right]_{r=r_\mathrm{v}}, \nonumber \\
v_{r\mathrm{v},\varphi}&=&\frac{1}{r_\mathrm{v}}\left[ \henb{v_r(r_\mathrm{v}, \varphi)}{\varphi}\right]_{\varphi=\varphi_\mathrm{v}}. \nonumber 
\end{eqnarray}
In addition, we define the radial and azimuthal convexities of the pressure function at the vortex center by 
\begin{eqnarray}
\Pi_{\mathrm{v},rr}&=&\left[ \henb{^2\Pi(r, \varphi_\mathrm{v})}{r^2}\right]_{r=r_\mathrm{v}}, \nonumber \\
\Pi_{\mathrm{v},\varphi \varphi}&=&\frac{1}{r_\mathrm{v}^2}\left[ \henb{^2\Pi(r_\mathrm{v}, \varphi)}{\varphi^2}\right]_{\varphi=\varphi_\mathrm{v}}. \nonumber 
\end{eqnarray}
The velocity gradients and the convexities of the pressure function are used to compare the RWI vortices with the analytic solutions of steady vortices (see Section 4.2.1). 
The velocity profiles gradually deviate from the straight lines with distance from the vortex center and finally have two extrema. 
At these extrema, the values of $|\delta v_\varphi |$ and $|v_r|$ are about two thirds times as large as $|(r-r_\mathrm{v})\delta v_{\varphi \mathrm{v}, r}|$ and $|r_\mathrm{v}(\varphi -\varphi_\mathrm{v})v_{r\mathrm{v}, \varphi}|$, respectively. 
We define the radial and azimuthal half widths $\Delta r$ and $r_\mathrm{v} \Delta \varphi$ by the half of the distance between the two extrema of $\delta v_\varphi$ and $v_r$. 
%check 2

\begin{figure*}
\epsscale{0.92}
\plotone{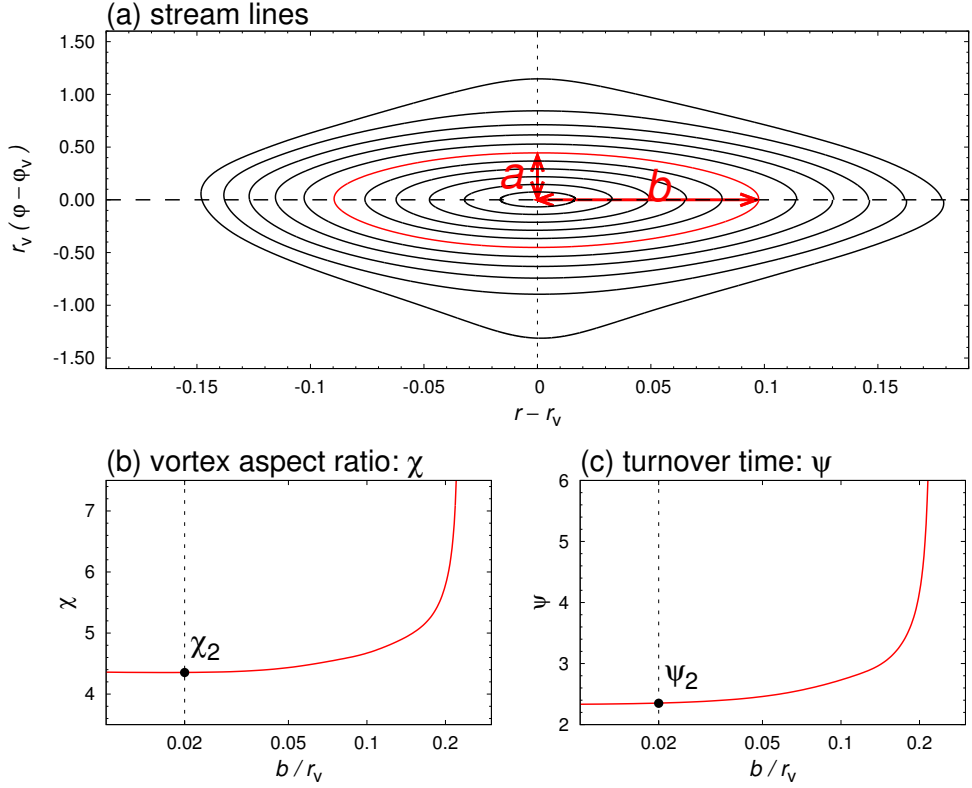}
\vspace{-1.2cm}
\caption{Panel (a) shows streamlines around the vortex center in the ''h10w3g1'' run at $\tau \ =\ 100$. Panels (b) and (c) show the profiles of $\chi$ and $\psi$ as functions of $b/r_\mathrm{v}$, respectively.}
\label{fig:stream}
\end{figure*}

The vortex aspect ratio and the turnover time in the vortex are important physical quantities of the vortex \citep[e.g.,][]{1981JPSJ..50..3517}. 
In this paper, we measure these quantities using streamlines. 
As shown in Panel (a) of Figure \ref{fig:stream}, the streamlines around the vortex center look like closed loops, indicating that the flow is in a quasi-stationary state. 
The streamlines are almost elliptic in the $(r\ -\ r_\mathrm{v})$--$r_\mathrm{v}(\varphi \ - \varphi_\mathrm{v})$ plane and the semi-minor axes of them are aligned to the radial direction. 
We measure the semi-minor axis $b$ (the radial direction) and the semi-major axis $a$ (the azimuthal direction) for each streamline. 
We define the aspect ratio of each streamline by $\chi \ \equiv \ a/b$. 
We also measure the turnover time of each streamline normalized by $2\pi/\Omega_\mathrm{v}$, $\psi$: 
\begin{equation}
\psi \equiv \frac{\Omega_\mathrm{v}}{2\pi} \oint_\mathcal{L} \frac{\mathrm{d} \ell}{\tilde{v}}, 
\end{equation}
where $\Omega_\mathrm{v} \ \equiv \ \Omega(r_\mathrm{v}) \ =\  \Omega_\mathrm{K}(r_\mathrm{v})$, $\mathcal{L}$ denotes the integration along the streamline, $\mathrm{d} \ell$ is the line element along the streamline, and $\tilde{v} \ = \ \sqrt{v_r^2+(v_\varphi - r\Omega_\mathrm{v})^2}$ is the magnitude of the velocity field in the rotating frame with the vortex center. 
Panels (b) and (c) of Figure \ref{fig:stream} show the profiles of $\chi$ and $\psi$ as functions of the normalized distance from the vortex center, $b/r_\mathrm{v}$. 
Both $\chi$ and $\psi$ are almost constant around the vortex center. 
At $b~\geq~0.1~r_\mathrm{v}$, these quantities are no longer constant and increase rapidly. 
Here, $0.1~r_\mathrm{v}$ is close to one disk scale height at the vortex center. 
We define $\chi_2$ and $\psi_2$ by $\chi$ and  $\psi$ at $b/r_\mathrm{v}\ =\ 0.02$ as representative vortex aspect ratio and normalized turnover time in the vortex, respectively. 
%check 2

\begin{table*}
\begin{center}
\caption{List of Physical Quantities and Measurements about the RWI Vortex.}
\label{tbl:lpv}
\renewcommand{\arraystretch}{1.0}
\begin{tabular}{lp{12cm}}
\tableline \tableline
Measurement&Meaning \\
\hline 
\hline
$r_\mathrm{v}$&The distance of the vortex center from the central star. \\
$\varphi_\mathrm{v}$&The azimuthal angle at the vortex center. \\
$\Omega_\mathrm{v} \ \equiv \ \sqrt{GM/r_\mathrm{v}^3}$& The angular velocity at the vortex center. \\
$\Sigma_\mathrm{v}$&The surface density at the vortex center. \\
$\delta v_\varphi$& The relative rotation velocity from the Keplerian velocity at the vortex center. \\
$\delta v_{\varphi \mathrm{v},r}$&The radial gradient of $\delta v_\varphi$ at the vortex center. \\
$v_{r\mathrm{v},\varphi}$&The azimuthal gradient of $v_r$ at the vortex center. \\
$\Pi_{\mathrm{v},rr}$&The radial convexity of the pressure function at the vortex center. \\
$\Pi_{\mathrm{v},\varphi \varphi}$&The azimuthal convexity of the pressure function at the vortex center. \\
$\Delta r$&The radial half width of the vortex. \\
$r_\mathrm{v}\Delta \varphi$&The azimuthal half width of the vortex. \\
$\chi_2$&The vortex aspect ratio in the vicinity of the vortex center. \\
$\psi_{2}$&The turnover time normalized in the vicinity of the vortex center. \\
$q_\mathrm{v}$&The vortensity at the vortex center. \\
$\xi$&The distance which the vortex moves in the $-r$ direction fro a unit time. \\
$\tau_\mathrm{mig}$& The vortex migration timescale. \\
$\chi_{2, \mathrm{ini}}$&The value of $\chi_2$ just after the RWI vortex formation. \\
$\tau_\chi$& The decrease timescale of the vortex aspect ratio. \\
\tableline \tableline
\end{tabular}
\end{center}
\end{table*}
%check2

\begin{figure*}
\epsscale{1.0}
\plotone{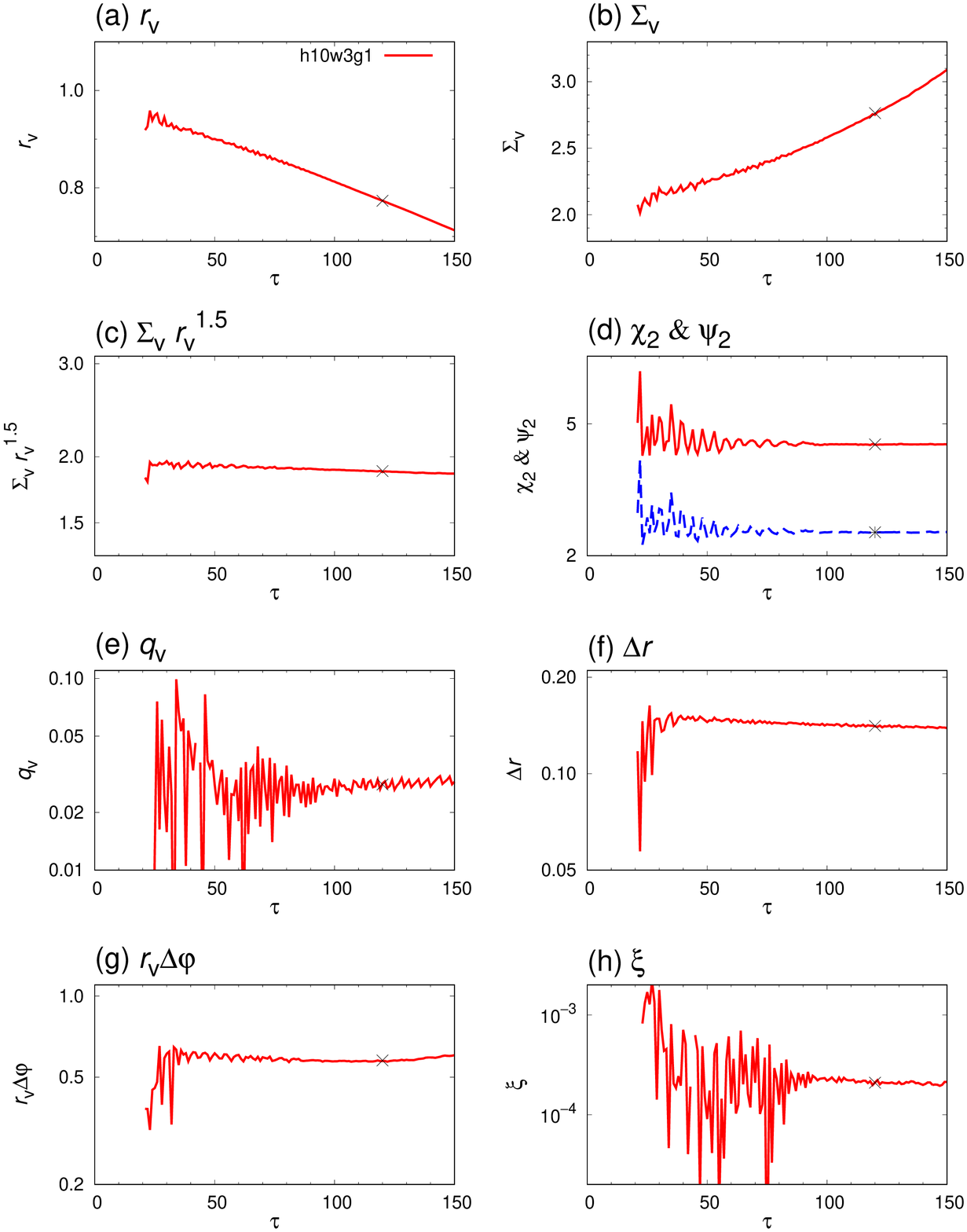}
\vspace{-1.1cm}
\caption{The time evolution of the RWI vortex in the ''h10w3g1'' run. Each panel shows the time evolution of (a) $r_\mathrm{v}$, (b) $\Sigma_\mathrm{v}$, (c) $\chi_2$ (red) and $\psi_2$ (blue), (d) $q_{v}$, (e) $\Delta r$, and (f) $r_\mathrm{v}\Delta \varphi$. The cross points represent the time-averaged values over 40 orbits at $\tau \ =\ 120$.}
\label{fig:uzu2}
\end{figure*}
%check2

\subsubsection{Measurements about the RWI Vortex}

We terminate all our main calculations after the disks have had a few hundred orbits at $r\ =\ r_\mathrm{n}$ since the RWI vortex formation. 
Here, we provide the way to measure the properties of the RWI vortices in our calculations. 
%check 2

From Panels (a) and (b) of Figures \ref{fig:uzu2}, the RWI vortices migrate toward the central star and their surface densities increase. 
At that time, the vortices keep $\Sigma_\mathrm{v} r_\mathrm{v}^{1.5}$ almost constant during the vortex migration as seen in Panel (c) of Figures \ref{fig:uzu2}. 
From Panels (d)--(g) of Figure \ref{fig:uzu2}, the vortex aspect ratio $\chi_2$, the vortex turnover time $\psi_2$, the vortensity at the vortex center $q_\mathrm{v} \ \equiv \ [\mathrm{rot}{\bf v}](r_\mathrm{v}, \ \varphi_\mathrm{v}) / \Sigma_\mathrm{v}$, the radial half width of the vortex $\Delta r$, and the azimuthal half width of the vortex $r_\mathrm{v}\Delta \varphi$ are approximately constant. 
To investigate the migration speed of the RWI vortex, we define a physical quantity $\xi$ by 
\begin{equation}
\xi \equiv -\frac{1}{\Omega_\mathrm{v}} \diff{r_\mathrm{v}}{t} = \frac{-1}{2.5\sqrt{GM}}\diff{r_\mathrm{v}^{2.5}}{t}, 
\end{equation}
where $\Omega_\mathrm{v}\ =\ \sqrt{GM/r_\mathrm{v}^3}$. 
The value of $\xi$ shows the distance which the vortex moves in the $-r$ direction for a unit time. 
We define the timescale of the vortex migration $\tau_\mathrm{mig}$ by $\tau_\mathrm{mig}\ \equiv \ \Omega_\mathrm{n}r_\mathrm{v}/(2\pi \xi \Omega_\mathrm{v})$. 
Panel (h) of Figure \ref{fig:uzu2} shows that $\xi$ is also almost constant. 
In this paper, we use the time-averaged values of these physical quantities over 40 orbits in each run as the measurements of the RWI vortex. 
We summarize the physical quantities and measurements about the RWI vortex in Table \ref{tbl:lpv}. 
%check 2

The RWI vortex is quasi-stationary but is not completely stationary. 
In a longer timescale than a thousand orbits, the values of  $\Sigma_\mathrm{v} r_\mathrm{v}^{1.5}$, $q_\mathrm{v}$, and $\Delta r$ are still almost constant, but the values of $\chi_2$, $\psi_2$, $r_\mathrm{v}\Delta \varphi$, and $\xi$ are not. 
We will discuss the applicability of our results to the long-term evolution in Section 4.3.2. 
However, we speculate that the long-term evolution is due to numerical viscosity in our calculations. 
Since viscosity seems to be generally important for the long-term evolution of the RWI vortices, the detailed investigation of the long-term evolution falls outside the scope of this paper. 
%check 2

\begin{figure*}
\epsscale{1.0}
\plotone{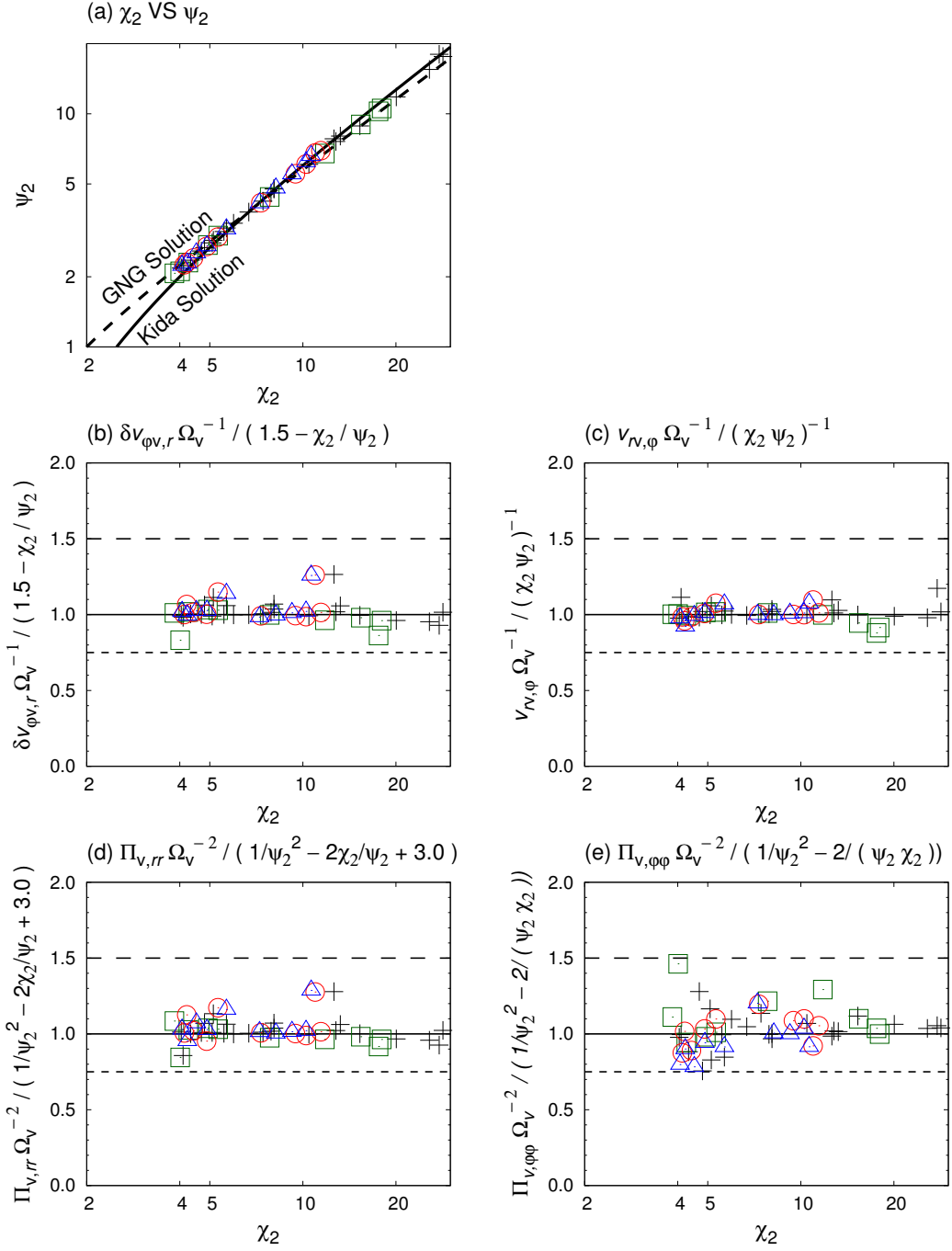}
\vspace{-1.1cm}
\caption{Comparing the RWI vortices with the analytic steady solutions of vortices. The time-averaged values of (a) $\psi_2$, (b) $\delta v_{\varphi \mathrm{v},r}\Omega^{-1}/(1.5-\chi_2/\psi_2)$, (c) $v_{r \mathrm{v},\varphi}\Omega_\mathrm{v}^{-1}/(\chi_2 \psi_2)^{-1}$, (d) $\Pi_{\mathrm{v},rr}\Omega_\mathrm{v}^{-2}/(1/\psi_2^2-2\chi_2/\psi_2+3.0)$, and (e) $\Pi_{\mathrm{v},\varphi \varphi}\Omega_\mathrm{v}^{-2}/(1/\psi_2^2-2/(\psi_2\chi_2))$ against the time-averaged $\chi_2$ are shown with the green squares ($h\ =\ 0.05$), the black cross points ($h\ =\ 0.1$), the red circles ($h\ =\ 0.15$), and the blue triangles ($h\ =\ 0.2$). The Kida vortex lies on the solid line and the GNG solution lies on the dashed line in Panel (a).}
\label{fig:steady}
\end{figure*}
%check2

\begin{figure*}
\epsscale{1.0}
\plotone{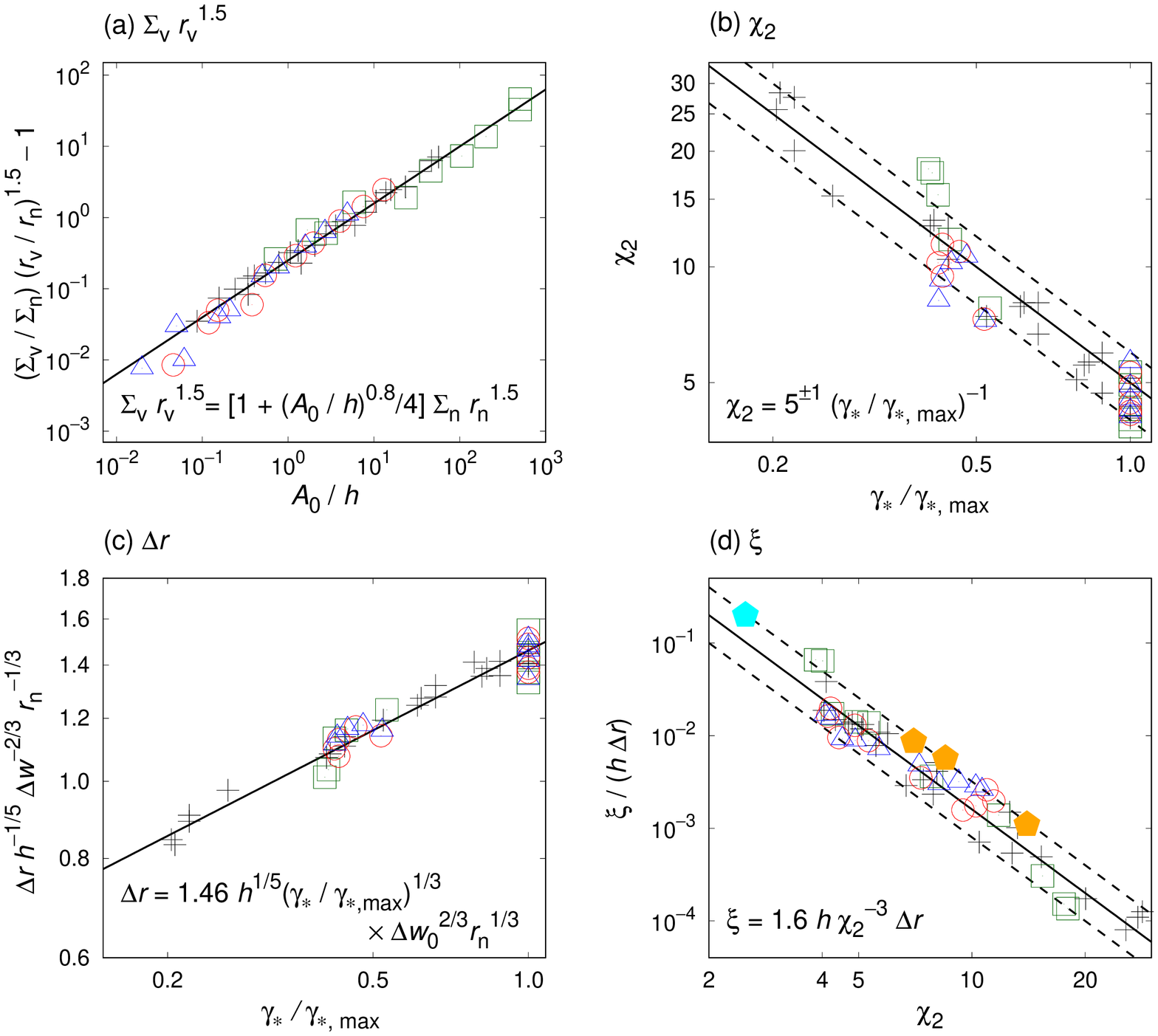}
\vspace{-0.6cm}
\caption{The values of (a) $\Sigma_\mathrm{v}r_\mathrm{v}^{1.5}$ against $\mathcal{A}_0/h$, (b) $\chi_{2, \mathrm{ini}}$ and (c) $\Delta r$ against $\gamma_\ast/\gamma_{\ast,\mathrm{max}}$, and (d) $\xi/(h\Delta r)$ against $\chi_2$ are shown with the green squares ($h\ =\ 0.05$), the black cross points ($h\ =\ 0.1$), the red circles ($h\ =\ 0.15$), and the blue triangles ($h\ =\ 0.2$). 
In Panel (a), the solid line shows $\Sigma_\mathrm{v}r_\mathrm{v}^{1.5}\ =\ [1+(\mathcal{A}_0/h)^{0.8}/4]\Sigma_\mathrm{n} r_\mathrm{n}^{1.5}$. 
In Panel (b), the solid line shows $\chi_{2} \ =\ 5(\gamma_\ast/\gamma_{\ast,\mathrm{max}})^{-1}$ and the dashed lines show $\chi_{2} \ =\ 4(\gamma_\ast/\gamma_{\ast,\mathrm{max}})^{-1}$ and $\chi_{2} \ =\ 6(\gamma_\ast/\gamma_{\ast,\mathrm{max}})^{-1}$ respectively. 
In Panel (c), the solid line shows $\Delta r\ =\ 1.46 h^{1/5}(\gamma_\ast/\gamma_{\ast,\mathrm{max}})^{1/3}\Delta w_0^{2/3}r_\mathrm{n}$. 
In Panel (d), the solid line shows $\xi\ =\ 1.6 h \chi_2^{-3}\Delta r$ and the dashed lines are twice and half of the solid line, respectively. 
The pentagons show the values derived from the results of \citet{2010ApJ...725..146P} (cyan) and \citet{2013A&A...559A..30R} (orange). }
\label{fig:empirical}
\end{figure*}
%check2

\subsubsection{Tracer Particle Analysis}
In order to obtain more detailed information of the RWI vortex, we perform a tracer particle analysis. 
We calculate path lines of fluid particles for 30 orbits in the ``h10w3g1'' run. 
$240\ \times \ 360$ tracer particles are initially distributed uniformly within the range of $0.6 \ \leq \ r\ \leq \ 1.2$ and $-\pi \ \leq \ \varphi \ \leq \ \pi$ at $\tau \ =\ 55,\ 60,\ 65,\ 70,\ 75,\ 80$, $85$. 
%check 2

\subsection{Results} \label{sec:result}

\subsubsection{Comparison with the Analytic Steady Vortices}
We compare the RWI vortices with known analytic solutions of steady vortices (the Kida and GNG solutions; see Appendix A in detail). 
The steady vortices satisfy 
\begin{eqnarray}
\delta v_{\varphi \mathrm{v},r} &=& (1.5-\chi_2/\psi_2)\Omega_\mathrm{v}, \label{eq:stap1} \\
v_{r\mathrm{v},\varphi} &=& \Omega_\mathrm{v} / (\chi_2 \psi_2) \label{eq:stap2}, \label{eq:stap2}\\
\Pi_{\mathrm{v},rr} &=& (1/\psi_2^2-2\chi_2/\psi_2+3.0)\Omega_\mathrm{v}^2, \label{eq:stap3} \\
\Pi_{\mathrm{v},\varphi \varphi} &=& [1/\psi_2^2-2/(\psi_2\chi_2)]\Omega_\mathrm{v}^2. \label{eq:stap4} 
\end{eqnarray}
Equations (\ref{eq:stap1})--(\ref{eq:stap4}) correspond to equations (\ref{eq:kvye})--(\ref{eq:kvxe}) and (\ref{eq:pfxx})--(\ref{eq:pfyy}). 
In the case of the Kida solution, the vortex aspect ratio $\chi$ and the vortex turnover time $\psi$ for the steady vortices are related as 
\begin{equation}
(\chi -1) = \mathfrak{S} \psi. \label{eq:Kida}
\end{equation}
The GNG solution \citep{1987MNRAS.225..695G} gives another relation: 
\begin{equation}
(\chi^2-1)=2\mathfrak{S}\psi^2. \label{eq:GNG}
\end{equation}
%check 2

As shown in Panel (a) of Figure \ref{fig:steady}, $\chi_2$ and $\psi_2$ of the RWI vortices reasonably satisfy equations equations (\ref{eq:Kida}) and (\ref{eq:GNG}). 
Panels (b)--(e) of Figure \ref{fig:steady} show that the RWI vortices satsify equations (\ref{eq:stap1})--(\ref{eq:stap4}) within a factor of 1.5. 
Therefore, the RWI vortices resemble the steady vortices in the Keplerian shear as shown in previous works \citep[e.g.,][]{2015A&A...579A.100S}. 
%check 2

\subsubsection{Empirical Relations  of the RWI Vortex}

\begin{figure}
\epsscale{1.1}
\plotone{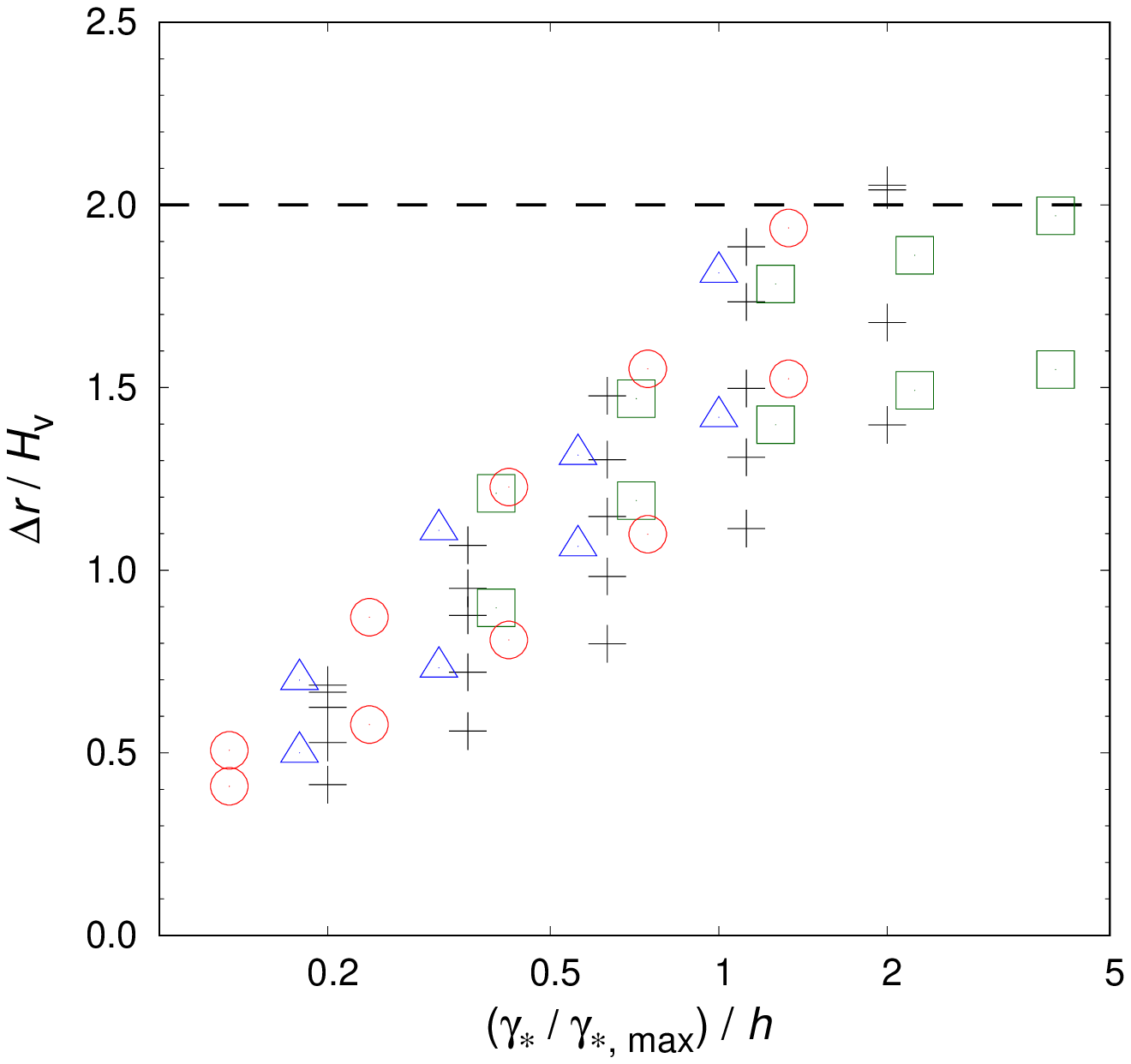}
\vspace{-0.4cm}
\caption{
The values of $\Delta r/H_\mathrm{v}$ against $(\gamma_\ast/\gamma_{\ast,\mathrm{max}})/h$ are shown with the green squares ($h\ =\ 0.05$), the black cross points ($h\ =\ 0.1$), the red circles ($h\ =\ 0.15$), and the blue triangles ($h\ =\ 0.2$). 
}
\label{fig:dr}
\end{figure}
%check2

We obtain some empirical relations between the RWI vortices and the initial bump structures so that they can help us predict the non-linear outcomes from initial structures. 
%check 2

First, we consider the surface density at the vortex center, $\Sigma_\mathrm{v}$. 
Compiling all the calculations with the different initial conditions, we find the empirical relations between $\Sigma_\mathrm{v} r_\mathrm{v}^{1.5}$ and the parameters of the initial bumps as (see Panel (a) of Figure \ref{fig:empirical}) 
\begin{equation}
\Sigma_\mathrm{v} r_\mathrm{v}^{1.5} \approx \left[ 1+\frac{(\mathcal{A}_0/h)^{0.8}}{4} \right] \Sigma_\mathrm{n} r_\mathrm{n}^{1.5}. \label{eq:sigest}
\end{equation}

Next, we consider the vortex aspect ratio $\chi_2$. 
Qualitatively, the vortex aspect ratio $\chi_2$ is small for the small growth rate cases, and vice versa. 
As shown in Panel (b) of Figure \ref{fig:empirical}, we find that $\chi_2$ and $\gamma_\ast / \gamma_{\ast, \mathrm{max}}$ are related by 
\begin{equation}
\chi_2 \approx 5^{\pm 1}\left( \gamma_\ast / \gamma_{\ast,\mathrm{max}} \right)^{-1}.  \label{eq:asest}
\end{equation} 
Using equation (\ref{eq:asest}), we can estimate the aspect ratio of the RWI vortex from the initial bump structures. 
Once we know the aspect ratio, it is possible to estimate the vortex turnover time, and the velocity gradients and the convexities of the pressure function at the vortex center using the analytic vortex solutions. 
%check 2

We turn our attention to the radial vortex size. 
We find that $\Delta r$ is related to the initial bump parameters by (see Panel (b) of Figure \ref{fig:empirical}) 
\begin{equation}
\Delta r \approx 1.46 h^{1/5}\left( \gamma_\ast / \gamma_{\ast, \mathrm{max}} \right)^{1/3}\Delta w_0^{2/3}r_\mathrm{n}^{1/3}. \label{eq:drest}
\end{equation}
We also compare $\Delta r$ with the disk scale height at the vortex center $H_\mathrm{v}\ \equiv \ h(\Sigma_\mathrm{v}/\Sigma_\mathrm{n})^{(\Gamma -1)/2} (r_\mathrm{v}/r_\mathrm{n})^{1.5}r_\mathrm{n}$. 
Figure \ref{fig:dr} indicates that a maximum value of $\Delta r / H_\mathrm{v}$ is about $2$. 
This maximum value of the radial vortex size is consistent with the values reported by previous works \citep[e.g.,][]{2001ApJ...551..874L, 2015A&A...579A.100S}. 
%check 2

Finally, we consider the vortex migration speed. 
From Panel (d) of Figure \ref{fig:empirical}, we find that $\xi$ satisfies
\begin{equation}
\xi = 1.6 h \chi_2^{-3} \Delta r, \label{eq:migest}
\end{equation}
within a factor of about 2. 
Using equation (\ref{eq:migest}), the timescale of the vortex migration $\tau_\mathrm{mig}$ is obtained as
\begin{equation}
\tau_\mathrm{mig}\approx 2.1\times 10^3 \left(\frac{h}{0.1}\right)^{-1} \left( \frac{\chi_\mathrm{2}}{6} \right)^{3} \left(\frac{\Delta r}{0.1\, r_\mathrm{n}}\right)^{-1} \left( \frac{r_\mathrm{v}}{r_\mathrm{n}} \right)^{2.5}. 
\end{equation}
If $r_\mathrm{n}\ =\ 100~\mathrm{AU}$ and $M\ =\ M_\odot$, an orbital time at $r\ =\ r_\mathrm{n}$ is about 1000 yrs so that the timescale of the vortex migration is 
\begin{equation}
2.1\ \mathrm{Myr} \times \left(\frac{h}{0.1}\right)^{-1} \left( \frac{\chi_\mathrm{2}}{6} \right)^{3} \left(\frac{\Delta r}{10~\mathrm{AU}}\right)^{-1} \left( \frac{r_\mathrm{v}}{100~\mathrm{AU}} \right)^{2.5}. 
\end{equation}
Unless the vortex aspect ratio is small ($\chi_2 \ \sim \ 4$) and the vortex size is large ($\Delta r \ \sim \ 2 H_\mathrm{v}$), the vortex migration timescale is comparable to or longer than the lifetime of the protoplanetary disks ($\sim$ 1--10 Myr). 
In other words, the RWI vortex resulting from a narrow and/or weak initial surface density bump stays within the disk without suffering from the radial migration. 
%check 2

Using equations (\ref{eq:sigest})--(\ref{eq:migest}), we can estimate the properties of the RWI vortex (the surface density at the vortex center, the aspect ratio, the radial size, and the migration speed) from the initial conditions ($h$, $\Delta w_0$, and $\gamma_\ast / \gamma_{\ast , \mathrm{max}}$). 
We note that $\mathcal{A}_0$ and $\gamma_{\ast , \mathrm{max}}$ can be calculated from $h$, $\Delta w_0$ and $\gamma_\ast$ performing the linear stability analyses. 
%check 2

\begin{figure}
\epsscale{1.2}
\plotone{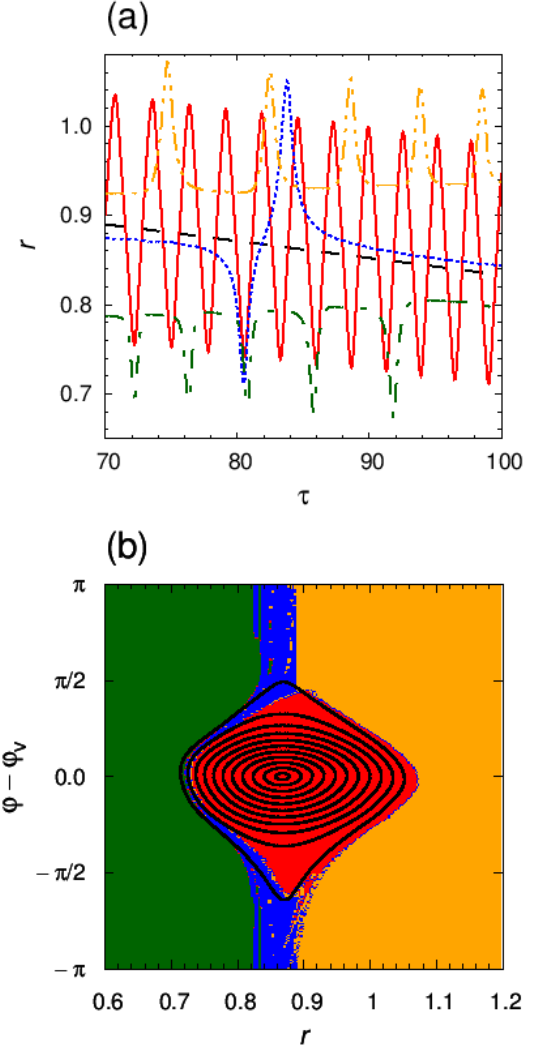}
\vspace{-1.1cm}
\caption{Categorization of the fluid particles into the four groups and distribution of the particle groups. The $r$ evolution of (a) the group I particle (the red solid line), (b) the group II particle (the blue dotted line), (c) the group III particle (the green one-dot chain line), and (d) the group IV particle (the orange two dots chain line) which depart from the starting points at $\tau \ =\ 70$ in the ``h10w3g1'' run is shown in Panel (a). The black dashed line shows the evolution of the vortex center. Panel (b) shows the distribution of the particle groups at $\tau \ =\ 70$ (red: the group I, blue: the group II, green: the group III, and orange: the group IV). The black lines show the streamlines in the vortex at $\tau \ =\ 70$. \label{fig:path1}}
\end{figure}
%check2

\begin{figure}
\epsscale{1.1}
\plotone{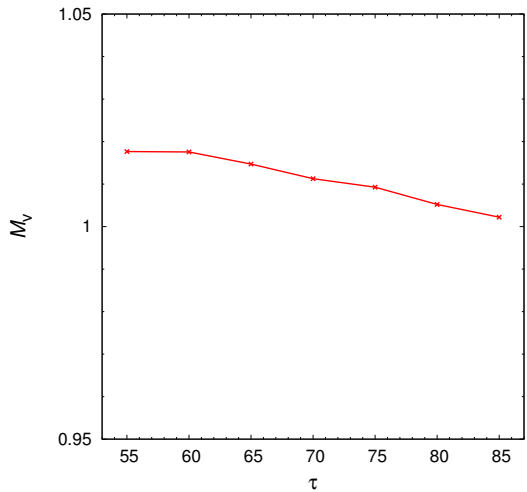}
\vspace{-0.5cm}
\caption{The time evolution of $M_\mathrm{v}$ in the ``h10w3g1'' model.}
\label{fig:path3}
\end{figure}
%check2

\subsubsection{The RWI Vortex from the Point of View of Tracer Particles}

We investigate the structure and evolution of the RWI vortex from the point of view of the tracer particles. 
We focus on the tracer particles initially distributed at $\tau \ =\ 70$ in the ``h10w3g1'' run. 
According to the $r$ evolution of those fluid particles (see Panel (a) of Figure \ref{fig:path1}), we categorize fluid particles into four groups (I, II, III, IV). 
The Group I represents the particles which compose the vortex at $\tau \ =\ 70$. 
All the group I particles remain in the vortex part during the 30 orbits. 
The fluid particles categorized into the group II are in the inner part of the disk initially and move to the outer part of the disk within the 30 orbits. 
The group III particles and the group IV particles remain in the inner part and in the outer part for the 30 orbits, respectively. 
The vortex captures a few particles which reside in the inner region, but such particles escape from the vortex to the outer part after only a few turnover motions and are categorized into the group II. 
We find that no particle moves inward (the outer part $\rightarrow$ the inner part, the outer part $\rightarrow$ the vortex part, and the vortex part $\rightarrow$ the inner part) during the 30 orbits. 
We show the distribution of the particle groups at $\tau \ =\ 70$ in Panel (b) of Figure \ref{fig:path1}. 
The tracer particles in the other calculations are also categorized in the same way. 
%check 2

We calculate the total mass of the group I particles, $M_\mathrm{v}$, which represents the vortex mass. 
From Figure \ref{fig:path3}, $M_\mathrm{v}$ is approximately constant within 1\% over 30 orbits in the ``h10w3g1'' run. 
We, therefore, expect that, in the course of the inward migration, the RWI vortex carries most of the fluid particles originating from the initial location where the vortex is formed, down to the inner radii. 
In this sense, the RWI vortex can be considered as a physical entity like a large fluid particle. 
%check 2

\subsection{Discussions}

\begin{figure}
\epsscale{1.2}
\plotone{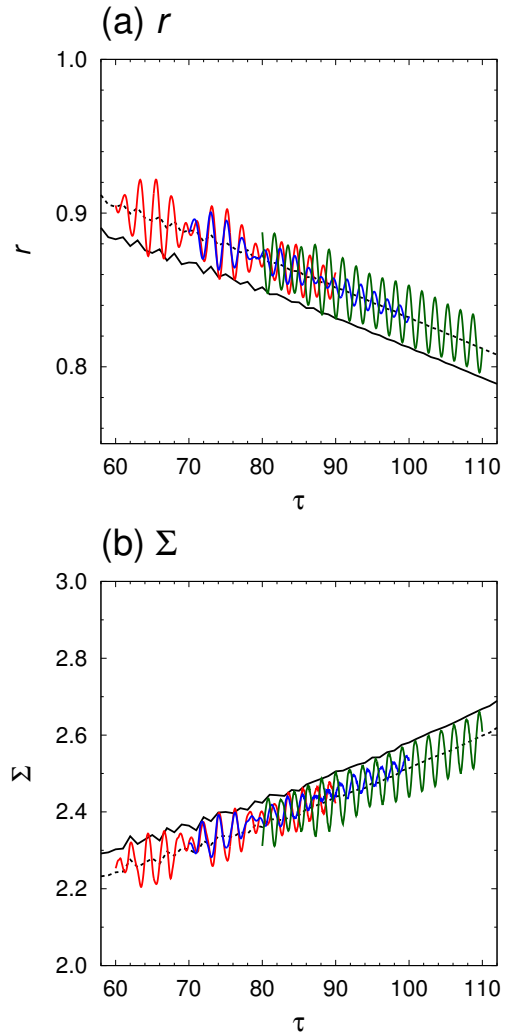}
\vspace{-1.3cm}
\caption{The $r$ and $\Sigma$ evolution of the fluid particles which depart from the starting points near the vortex center at $\tau \ =\ 60$ (the red line), $70$ (the blue line), and $80$ (the green line) are shown in Panels (a) and (b), respectively. 
In Panel (a), the black solid line shows the evolution of $r_\mathrm{v}$ and the black dotted line show the evolution of $1.022r_\mathrm{v}$. In Panel (b), the black solid line shows the evolution of $\Sigma_\mathrm{v}$ and the black dotted line shows the evolution of $0.974\Sigma_\mathrm{v}$. \label{fig:path2}}
\end{figure}
%check2

\subsubsection{Definition of the Vortex Center}

Our definition of the vortex center is based on the velocity field. 
The vortex center defined from the velocity field corresponds to the center of the streamlines in the vortex. 
Therefore, we call the vortex center obtained from our definition ``streamline vortex center''. 
However, There is another way to obtain the vortex center using tracer particles. 
To focus on the time evolution of the tracer particles close to the vortex center, we use three particles which are located near the vortex center at $\tau \ =\ 60,\ 70$, and $80$ in the ``h10w3g1'' run. 
From Panel (a) of Figure \ref{fig:path2}, those particles migrate inward with small oscillations. 
Here, we can define the center of the oscillations by ``particle vortex center''. 
%check 2

The distance of the particle vortex center from the central star is larger than $r_v$ by about 2.2\%. 
We consider that this discrepancy originates from the difference between the position of the surface density peak and that of the vortensity minimum. 
Since the dynamical equilibrium is achieved in the vicinity of the surface density peak, the flow should be stagnant there. 
Therefore, the streamline vortex center is very close to the surface density peak. 
On the other hand, the particle vortex center is very close to the vortensity minimum due to the vortensity conservation law. 
%check 2

However, the migration speed of both vortex centers almost corresponds as shown in Panel (a) of Figure \ref{fig:path2}. 
We can also confirm this fact in Panel (b) of Figure \ref{fig:path2}. 
The time-averaged surface densities of the three particles are smaller than $\Sigma_\mathrm{v}$ by about 2.6\%, but the growth rates of the surface densities of these tracer particles are almost equivalent to that of $\Sigma_\mathrm{v}$. 
Therefore, the choice of the vortex center has no effect on our results. 
%check 2

\begin{figure*}
\plotone{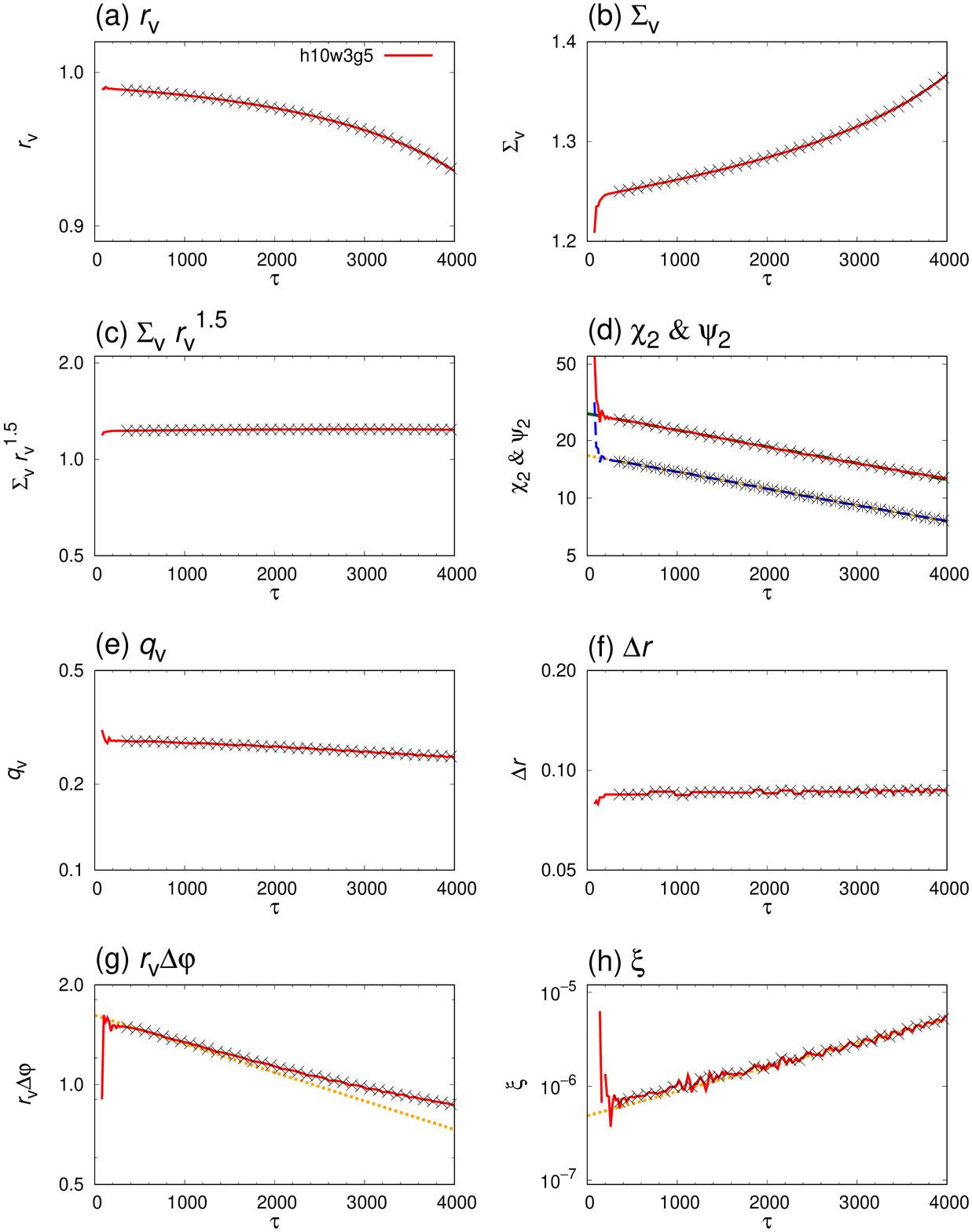}
\vspace{-1.4cm}
\caption{Similar to Figure \ref{fig:uzu2}, but the long-term calculation of the ''h10w3g5'' model. The cross points show the time-averaged values over 40 orbits every 100 orbits from $\tau\ =\ 360$. The green dashed line in Panel (d) shows the best fit of the exponential decreases of $\chi_2$ with $\chi_{2, \mathrm{ini}}\ =\ 27.6$ and $\tau_\chi \ =\ 5036$. 
The orange dotted lines show $\psi_2\ =\ 16.7\ \exp(-\tau/\tau_\chi)$ in Panel (d), $r_\mathrm{v} \Delta \varphi \ =\ 1.62 r_\mathrm{n}\ \exp(-\tau/\tau_\chi)$ in Panel (g), and $\xi \ =\ 4.85\times 10^{-7} r_\mathrm{n}\ \exp(3\tau/\tau_\chi)$ in Panel (h). }
\label{fig:uzu3}
\end{figure*}
%check2

\subsubsection{Applicability of Empirical Formulae to the Long Term Evolution}

In cases with a large linear growth rate, the RWI vortices migrate too fast to survive for a thousand orbits. 
On the other hand, the vortex migration is slow enough to survive for a thousand orbits in cases with a small linear growth rate. 
In such a long timescale, not all the physical quantities that are approximately constant in a short timescale are almost constant. 
In this section, we check the applicability of the empirical formulae obtained in Section 4.2.2 to the long-term evolution. 
%check 2

Here, we regard the "h10w3g5" run ($h\ =\ 0.1,\ \Delta w_0\ =\ 0.0632r_\mathrm{n}$,\ $\gamma_\ast / \Omega_\mathrm{n} \ =\ 0.1$, $ \mathcal{A}_0\ =\ 0.142$, and $m_\ast\ =\ 2$) as a representative case with a small linear growth rate. 
From Figure \ref{fig:uzu3}, $\Sigma_\mathrm{v} r_\mathrm{v}^{1.5}$, $q_\mathrm{v}$, and $\Delta r$ are still almost constant, but $\chi_2$, $\psi_2$, and $r_\mathrm{v}\Delta \varphi$ decrease and $\xi$ increases in the long-term calculation of the ``h10w3g5'' model. 
Panel (d) of Figure \ref{fig:uzu3} shows that the decrease of $\chi_2$ is exponential-like following 
\begin{equation}
\chi_2 = \chi_{2, \mathrm{ini}} \exp(-\tau/\tau_\chi), 
\end{equation} 
where $\chi_{2,\mathrm{ini}}\ \approx \ 27.6$ is the vortex aspect ratio just after the RWI vortex formation and $\tau_\chi \ \approx \ 5040$ is the decrease timescale of the vortex aspect ratio. 
Panels (d) and (g) of Figure \ref{fig:uzu3} show that $\psi_2$ and $r_\mathrm{v}\Delta \varphi$ also exponentially decrease on a similar timescale to $\tau_\chi$ and on a slightly longer timescale, respectively. 
From panel Panel (h) of Figure \ref{fig:uzu3}, $\xi$ exponentially increases on a similar timescale to $3 \tau_\chi$. 
We consider that $\tau_\chi$ shows the timescale of the long-term evolution. 
%check 2

We calculate the time-averaged measurements of the RWI vortex over 40 orbits every 100 orbits. 
Since $\Sigma_\mathrm{v} r_\mathrm{v}^{1.5}$ and $\Delta r$ are still approximately constant, equations (\ref{eq:sigest})  and (\ref{eq:drest}) are satisfied. 
We also confirm that the RWI vortex always resembles the steady vortices and satisfies equation (\ref{eq:migest}) even though $\chi_2$, $\psi_2$, and $\xi$ vary. 
However, the value of $\chi_2$ goes away from the value obtained by equation (\ref{eq:asest}). 
We observe these trends in the long-term calculations of the other small growth rate cases. 
Therefore, we conclude that the empirical formulae except for equation (\ref{eq:asest}) are applicable even to the long-term evolution. 
On the other hand, equation (\ref{eq:asest}) is applicable only for first a few hundred orbits after the RWI vortex formation. 
%check 2

Since $\Delta r$ is approximately constant, the shrink of the RWI vortex in the azimuthal direction can explain the long-term evolution of the vortex aspect ratio. 
In all the models without a large linear growth rate, $\tau_\chi$ is longer than 1000. 
For the cases with a large linear growth rate, it is impossible to measure $\tau_\chi$ due to the fast vortex migration. 
However, $\tau_\chi$ is expected to be also large compared to the migration timescale because the structure of the RWI vortex is almost stationary in the migration timescale. 
The values of $\tau_\chi$ are shown in Appendix C. 2. 
We also perform a long-term calculation of the ``h10w3g5'' model with a twice coarser resolution than that of the fiducial setup. 
We find that $\tau_\chi$ is several times smaller in the coarse calculation than in the fiducial calculation. 
This indicates that the numerical viscosity has a significant effect on the long-term evolution and the viscosity is important for the long-term evolution in general. 
Therefore, the detailed investigation of the long-term evolution is located outside the scope of this paper. 
%check 2

\subsubsection{Reason Why $\Sigma_\mathrm{v} r_\mathrm{v}^{1.5}$ is Approximately Constant}

As shown in Section 4.1.2 and Section 4.3.2, $\Sigma_\mathrm{v} r_\mathrm{v}^{1.5}$ is approximately constant as well as $q_\mathrm{v}$ in short and long timescales. 
In our calculations, the vortensity conservation law should be satisfied without taking the numerical viscosity into account. 
Almost the same tracer particles constitute the vortex center (see Section 4.2.3) so that the vortensity at the vortex center is approximately constant. 
In this section, we discuss the reason for the invariance of $\Sigma_\mathrm{v} r_\mathrm{v}^{1.5}$ using the vortensity conservation. 
%check 2

Here, we assume that the RWI vortex coincides to the Kida vortex. 
From the Kida solution, we obtain 
\begin{eqnarray}
\frac{\Sigma_\mathrm{v}}{\Omega_\mathrm{v}}q_\mathrm{v} &=& 2.0 - \frac{\chi_2}{\psi_2} - \frac{1}{\chi_2 \psi_2}, \nonumber \\
&=& \frac{1}{2} - \frac{3}{2} \left(\frac{2}{\chi_2-1} - \frac{1}{\chi_2} \right). \label{eq:qvcons}
\end{eqnarray}
For the large growth rate cases, the second term of equation (\ref{eq:qvcons}) is comparable to the first term because $\chi_2$ is small. 
In these cases, the vortex migration is fast and $\tau_\mathrm{mig}\ <\ \tau_\mathrm{\chi}$.  Therefore, we can safely assume that $\chi_2$ is constant over the vortex lifetime and $\Sigma_\mathrm{v}q_\mathrm{v}/\Omega_\mathrm{v}$ is constant. 
On the other hand, if linear growth rates are small, the vortex aspect ratio $\chi_2$ is large so that $\Sigma_\mathrm{v}q_\mathrm{v}/\Omega_\mathrm{v}$ is approximately constant at 1/2. 
Therefore, the values of $\Sigma_\mathrm{v}q_\mathrm{v}/\Omega_\mathrm{v}$ can be regarded as constant in all the cases. 
Due to the vortensity conservation law and the definition of the vortex center, $\Sigma_\mathrm{v}q_\mathrm{v}/\Omega_\mathrm{v}$ is proportional to $\Sigma_\mathrm{v}r_\mathrm{v}^{1.5}$. 
This is the reason why $\Sigma_\mathrm{v}r_\mathrm{v}^{1.5}$ is approximately constant in short and long timescales. 
%check 2

\begin{figure}
\epsscale{1.1}
\plotone{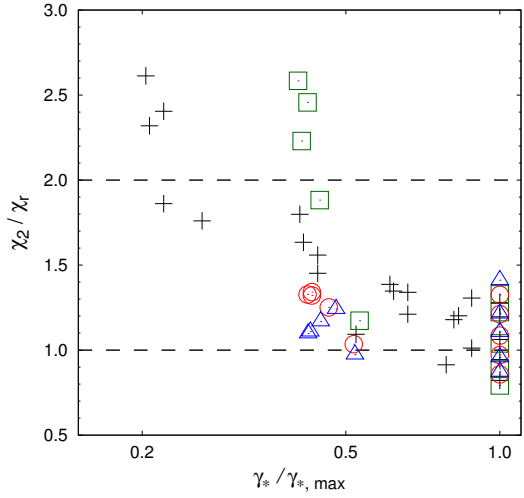}
\vspace{-0.5cm}
\caption{
The values of $\chi_2 / \chi_\mathrm{r}$ against $\gamma_\ast/\gamma_{\ast,\mathrm{max}}$ are shown with the green squares ($h\ =\ 0.05$), the black cross points ($h\ =\ 0.1$), the red circles ($h\ =\ 0.15$), and the blue triangles ($h\ =\ 0.2$). 
}
\label{fig:cr}
\end{figure}

\subsubsection{Comparison with Another Formula of Vortex Aspect Ratio}

In Section 4.2.2, we obtain the empirical formula of the vortex aspect ratio by equation (\ref{eq:asest}). 
\citet{2013A&A...559A..30R} derived another formula assuming the vorticity of the non-Keplerian motion normalized by that of the background shear flow, $\tilde{\omega}_z$, is steady at the peak of the initial bump and the vortex center. 
Here, we compare the two formulae. 
%check 2

When the profile of the initial surface density is given as a Gaussian bump, the value of $\tilde{\omega}_z$ at $r\ =\ r_\mathrm{n}$ for the initial conditions is calculated as 
\begin{eqnarray}
\tilde{\omega}_{z} &=& \left[ \diff{\{ r (v_{\varphi 0} - v_\mathrm{K})\} }{r} \, /\, \henb{\{ r (v_\mathrm{K} - r\Omega_\mathrm{n})\} }{r} \right]_{r=r_\mathrm{n}} \nonumber \\
&=& \frac{h^2\mathcal{A}_0(1+\mathcal{A}_0)^{\Gamma-2}}{3(\Delta w_0)^2}r_\mathrm{n}^2 \label{eq:ric1}. 
\end{eqnarray}
From the Kida solution, the RWI vortex should satisfy 
\begin{equation}
\tilde{\omega}_{z} = \frac{1+\chi}{\chi(\chi-1)}. \label{eq:ric2}
\end{equation}
\citet{2013A&A...559A..30R} estimated the vortex aspect ratio, which we denote by $\chi_\mathrm{r}$, from the balance between equations (\ref{eq:ric1})--(\ref{eq:ric2}). 
%check 2

From Figure \ref{fig:cr}, $\chi_{2}$ matches with $\chi_\mathrm{r}$ in the cases with a large linear growth rate. 
On the other hand, $\chi_{2}$ is larger than $\chi_\mathrm{r}$ by a factor of a few in the cases with a small linear growth rate. 
Therefore, we consider that the formula by \citet{2013A&A...559A..30R} is applicable and tested for the large growth rate cases and that our formula extends their work to the small growth rate cases. 
We note again that equation (\ref{eq:asest}) is applicable only for first a few hundred orbits after the RWI vortex formation (see Section 4.3.2). 
%check 2

\subsubsection{Generality of Empirical Formula of Vortex Migration Speed}

Now, we can estimate the migration speed of the RWI vortex using equation (\ref{eq:migest}). 
This empirical formula does not depend on the initial surface density profile explicitly. 
Here, we investigate whether the formula is generally applicable to vortices on disks or not. 
%check 2

\citet{2010ApJ...725..146P} reported the migration speed of the vortex which is formed by imposing a vorticity perturbation in their 2D disk for $h\ =\ 0.1$ at $r\ \approx \ r_\mathrm{n}$. 
The parameters of the vortex are ($\chi$, $\Delta r$, $\xi$) $=$ ($2.5$, $0.025\, r_\mathrm{n}$, $5.0\times 10^{-4} \, r_\mathrm{n}$). 
\citet{2013A&A...559A..30R} also measured the migration speed of three RWI vortices in their 3D calculations for $h\ \approx \ 0.0662$ at $r\ \approx \ r_\mathrm{n}$. 
The parameters of the three vortices are ($\chi$, $\Delta r$, $\xi$) $=$ ($7.0$, $0.0543\, r_\mathrm{n}$, $3.1\times 10^{-5} \, r_\mathrm{n}$), ($8.5$, $0.0531\, r_\mathrm{n}$, $2.0\times 10^{-5} \, r_\mathrm{n}$), and ($14.0$, $0.0413\, r_\mathrm{n}$, $3.0\times 10^{-6} \, r_\mathrm{n}$). 
We note that the numerical setups in \citet{2010ApJ...725..146P} and \citet{2013A&A...559A..30R} are somewhat different from our setup, where the initial surface density profiles have global radial gradients and the disks are assumed to be locally isothermal. 
In Panel (d) of Figure \ref{fig:empirical}, we plot the result of \citet{2010ApJ...725..146P} with a cyan pentagon and the results of \citet{2013A&A...559A..30R} with orange pentagons. 
We find that these points are almost on the line of $\xi \ = \ 3.2 h \chi_2^{-3} \Delta r$. 
%check 2

\citet{2010ApJ...725..146P} showed that the inward migration of vortices is faster if the global radial gradient of the surface density is negative and steeper. 
In our calculations, we assume that the radial profile of the surface density is globally flat. 
On the other hand, both \citet{2010ApJ...725..146P} and \citet{2013A&A...559A..30R} assumed that the radial profile is globally proportional to $r^{-1.5}$ so that $\xi$ is twice larger. 
In that sense, our results are qualitatively consistent with \citet{2010ApJ...725..146P}. 
In addition, equation (\ref{eq:migest}) is satisfied within a factor of two even in the calculations with such a surface density slope. 
This indicate that the dependence of the migration speed on the vortex structure and the disk aspect ratio seems to be universal. 
%check 2

According to \citet{2010ApJ...725..146P}, a pressure bump can prevent the vortex from migrating inward. 
In their calculations, the pressure bump is stronger and wider than the vortex. 
The axisymmetric components also have a pressure bump as shown in Figure \ref{fig:res_ave} in our calculations. 
However, it seems that the migration of the RWI vortices does occur even in the presence of the pressure bump. 
We consider that the pressure bump structures seen in our calculations are too weak and narrow to stop the vortex migration. 
One notable difference between \citet{2010ApJ...725..146P} and our work is that the structure of the pressure bump is determined consistently with the development of the RWI and the formation of the vortex. 
On the other hand, \citet{2010ApJ...725..146P} uses the parameterized model for the pressure bump without calculating its formation process. 
%check 2

In short, we consider that equation (\ref{eq:migest}) is broadly applicable to estimate the vortex migration speed regardless of the formation mechanism of the vortex in both 2D and 3D disks unless disks have very steep global slopes of the surface density or strong and wide pressure maxima. 
Further investigations are necessary to study quantitatively the effects of the global gradients of the surface density and the pressure maxima on the vortex migration. 
%check 2

\begin{figure*}
\plotone{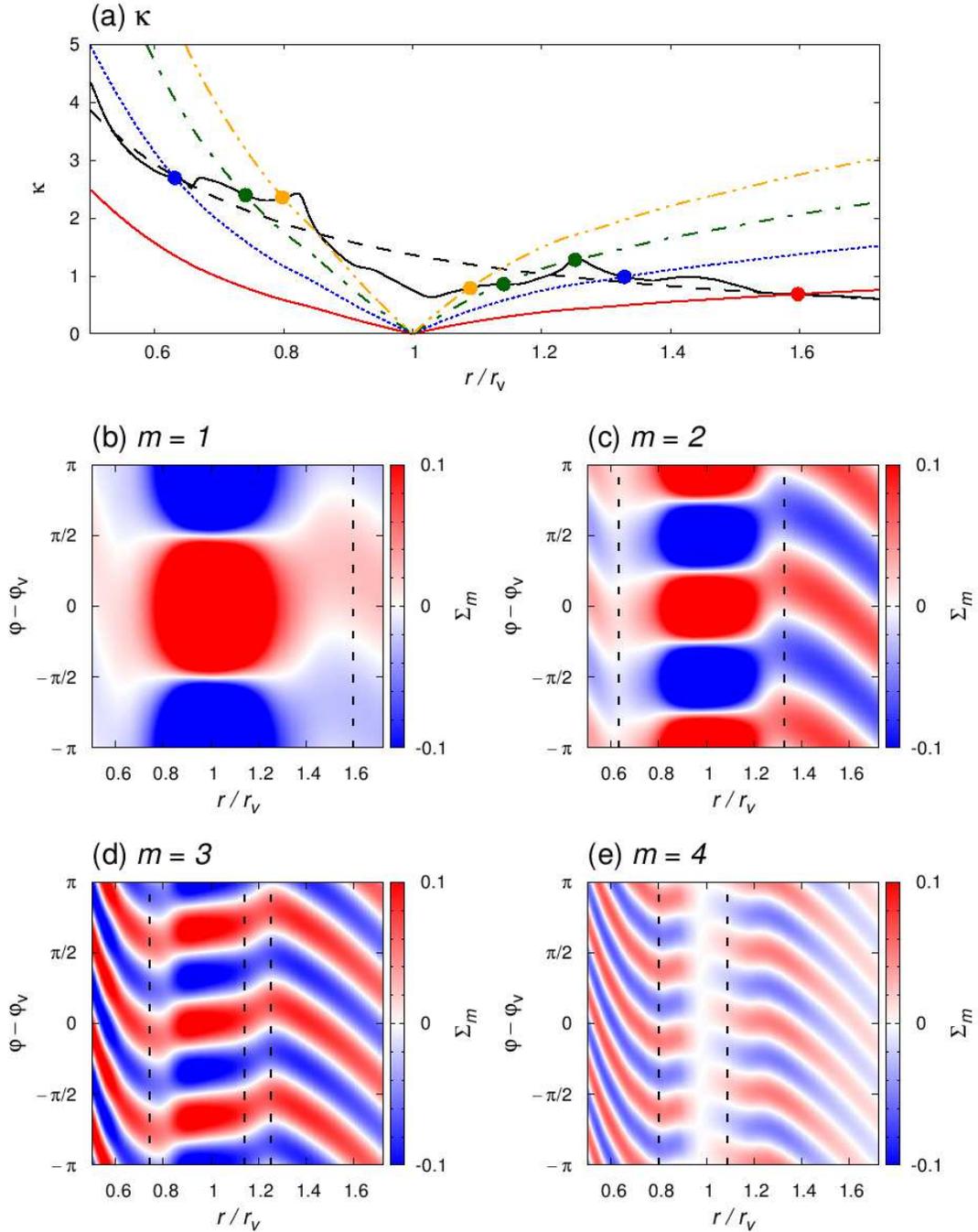}
\vspace{-1.5cm}
\caption{The Lindblad resonances and the surface density distribution for each $m$ mode. Panel (a) shows the epicyclic frequency $\kappa$ in the ``h10w3g1'' run at $\tau \ =\ 100$. In Panel (a), $m(\Omega_\mathrm{v}-v_\varphi/r)$ is shown with the red solid line ($m\ =\ 1$), the blue doted line ($m\ =\ 2$), the green one-dot chain line ($m\ =\ 3$), and the orange two dots chain line ($m\ =\ 4$) and $\Omega_K$ is shown with the black dashed line. Panels (b)--(e) show the 2D distribution of $\Sigma_m$ for $m\ =\ 1$--$4$, respectively. The points in Panel (a) and the vertical dashed lines in Panels (b)--(e) indicate the Lindblad resonances of each $m$ mode.}
\label{fig:mig2}
\end{figure*}

\subsubsection{Mechanism of Vortex Migration}
In this section, We discuss the mechanism of the vortex migration. 
During the vortex migration, the vortex loses the angular momentum via density waves \citep{2010ApJ...725..146P}. 
The velocity perturbations induced by the vortex motion excite density waves in a disk, which carry away negative (inner spirals) or positive (outer spirals) angular momentum, causing the vortex to migrate. 
The positions of the Lindblad resonances for the $m$ mode are located where the epicyclic frequency, $\kappa$, is the $m$ times of the angular velocity of the fluid element in the frame co-rotating with the vortex (see Panel (a) of Figure \ref{fig:mig2}): 
\begin{equation}
\kappa(r,\varphi_\mathrm{v}) = \pm m[\Omega_\mathrm{v} - v_\varphi(r,\varphi_\mathrm{v})/r]. 
\end{equation}
As can be seen in Panels (b)--(e) of Figure \ref{fig:mig2}, the density waves are indeed excited around the Lindblad resonances for each mode.
%check 2

The fluid particles of the group II (see Section 4.2.3) also contribute to the angular momentum exchange between the vortex and the disk. 
When they move from the inner part to the outer part, they gain the angular momentum from the vortex. 
It is analogous to the co-rotation torque exerted on a planet in the planet-disk interaction \citep{2001MNRAS.326..833B}. 
We calculate the variation of the total angular momentum of the group I particles, $\Delta J_1$, and that of group II particles, $\Delta J_2$, over the 30 orbits. 
We find $|\Delta J_2/\Delta J_1|\ \sim \ 0.2$, indicating that about 20\% of the total torque exerted on the vortex comes from the contribution of the fluid elements passing through the vortex region. 
The remaining about 80\% of the total torque is interpreted to originate from the density waves. 
We, however, find difficulty in measuring the total torque which comes from the density waves directly, because it is difficult to distinguish the density waves from the vortex motion and to precisely measure the angular momentum flux of the density waves due to numerical viscosity. 
In order to quantify the significance of each mechanism in more detail, we need to precisely measure both types of torque. 
%check 2

\subsubsection{Other Physical Effects}

In this paper, we consider the simplest disk model to keep the broad parameter search tractable. 
Various, potentially important, physical effects on the RWI, such as viscosity, 3D, self-gravity, and dust drag are not included. 
Here, we briefly discuss how these effects can affect the RWI vortex. 
%check 2

Viscosity has a significant effect on the lifetime of the RWI vortex \citep{2014MNRAS.437..575L}. 
For a long-term survival of the RWI vortex, very low viscosity ($\alpha \ \lesssim \ 10^{-4}$) is required, where $\alpha$ is the kinematic viscosity normalized by the sound speed and Kepler time \citep[the $\alpha$-parameter,][]{1973A&A....24..337S}. 
We expect that the circumstance with such low $\alpha$ is achieved within the MRI-dead zone \citep{1996ApJ...457..355G}. 
As discussed in Section 4.3.2, we speculate that viscosity is still important for the long-term evolution of the RWI vortex even for very low viscosity. 
The long-term behavior of the vortex needs further investigations with explicit viscosity prescription. 
%check 2

\citet{2009A&A...498....1L} reported that the effect of 3D can destroy vortices by the ellipsoidal instability. 
The analytic steady vortices are strongly unstable for $\chi \ \leq \ 4$ and weekly unstable for $\chi \ \gtrsim \ 6$ from the 3D linear stability analyses and the local numerical simulations in incompressible flow with Keplerian shear. 
For $\chi \ \lesssim \ 4$, the velocity field of the vortex violates the Rayleigh's condition and the vortex is destroyed. 
On the other hand, the vortex with $\chi \ \gtrsim \ 6$ is destroyed due to the resonance between the turnover motion of the vortex and the epicyclic motion of the disk. 
\citet{2013A&A...559A..30R} performed 3D compressible simulations of vortices formed by the RWI. 
In those simulations, the destruction of the vortices with $\chi \ \lesssim \ 4$ is verified. 
However, the vortices with $\chi \ \approx \ 7$ are not destroyed and survive for a sufficiently long time. 
Since our calculations are within the 2D framework, the RWI vortex does not suffer from the ellipsoidal instability. 
In all our calculations except for the ``h05w4g1'' and ``h05w5g1'' runs, $\chi_2$ is always larger than $4$. 
Even in the ``h05w4g1'' and ``h05w5g1'' runs, the vortex aspect ratio is approximately equal to 4. 
This is because we only consider the initial conditions that do not violate Rayleigh's criterion. 
In fact, we confirm that $\chi_2$ is smaller than 4 if the initial conditions violate the Rayleigh's condition, but we consider that such initial conditions are not realistic. 
Therefore, we expect that the evolution of the RWI vortices formed in our calculations almost never changes even in the 3D frameworks. 
%check 2

The effect of self-gravity prevents the onset of the RWI \citep{2013MNRAS.429..529L, 2016GApFD.110..274Y}. 
However, once the RWI vortex is formed, there are possibilities that self-gravity can help the vortex survive for a long time \citep{2018MNRAS.478..575L}. 
At that time, self-gravity can also have effects on the RWI evolution. 
%check 2

The effects of the dust particles on the gas flow are negligible in typical protoplanetary disks due to the low dust-to-gas mass ratio ($\sim \ 10^{-2}$). 
When a protoplanetary disk has a gas vortex, the vortex captures the dust particles and concentrates them at the vortex center \citep{1995A&A...295L...1B}. 
If sufficiently high dust-to-gas mass ratio is realized at the vortex center, the vortex is destroyed due to the gas-dust interaction \citep{2014ApJ...795L..39F, 2015MNRAS.450.4285C}. 
The motion of dust particles should be explored further as well as the hydrodynamics of the gas to understand how the RWI vortices act as the location where dust particles are accumulated. 
%check 2

We have fixed the value of the effective adiabatic index and have not
taken into account the baroclinicity and the global radial gradient of the initial surface density profile. 
In addition, the efficiency of disk cooling is also important on the RWI vortex \citep{2018MNRAS.tmp.1263P}. 
In order to investigate the applicability of the empirical relations obtained in this paper, a further parameter survey taking into account these other physical effects is needed. 
%check 2

\section{Summary}

We perform numerical simulations of the RWI in 2D, barotropic and hydrodynamic disks using the Athena++ code. 
As initial conditions, we consider axisymmetric disks with a Gaussian surface density bump. 
We have three parameters to characterize the initial bump: the dimensionless disk aspect ratio $h$, the radial half-widths $\Delta w_0$, the largest linear growth rate of the RWI $\gamma_\ast$. 
We vary these parameters in a wide parameter space and explore the non-linear evolution for 54 models. 
%check 2

First, we investigate the RWI evolution. 
Perturbations grow as expected by the linear stability analyses not only in the linear regime but also in the weakly non-linear regime. 
The axisymmetric component evolves as the RWI develops. 
When the axisymmetric component becomes marginally stable against the RWI for the most unstable azimuthal mode of the initial condition, the RWI saturation occurs and multiple vortices are formed in accordance with the most unstable azimuthal mode. 
After the RWI saturation, the vortices coalesce one after another. 
The axisymmetric component also approaches the stable configuration against the RWI during the vortex mergers. 
In the end, one quasi-stationary vortex (RWI vortex) remains when the axisymmetric component reaches the marginally stable configuration for the $m\ =\ 1$ mode. 
The regime with more than two vortices continues at most for a few orbits and the two vortices regime continues up to about 100 orbits. 
We conclude that it is difficult to observe the disks with multiple vortices originating from the RWI of one initial surface density bump except at outer disks. 
%check 2

Next, we turn our attention to the RWI vortex. 
We confirm that the RWI vortex almost corresponds to the analytic steady vortices on the Keplerian shear (the Kida solution and the GNG solution) as shown in previous works. 
Comparing the measurements of the RWI vortex with the initial conditions, we obtain empirical relations between the properties of the RWI vortices (the surface density at the vortex center: equation (\ref{eq:sigest}), aspect ratio: equation (\ref{eq:asest}), the radial size: equation (\ref{eq:drest}), and migration speed: equation (\ref{eq:migest})) and the initial conditions. 
The radial half-width of the RWI vortex is no larger than twice the disk scale height at the vortex center. 
Finally, we find that the RWI vortex can be considered as a physical entity like a large fluid particle from the tracer particle analysis. 
%check 2

Our results are not affected by the definition of the vortex center. 
Even if we take into account the long-term evolution of the RWI vortex, the empirical formulae except for equation (\ref{eq:asest}) are still applicable. 
On the other hand, equation (\ref{eq:asest}) is applicable only for first a few hundred orbits after the RWI vortex formation. 
We consider that viscosity is responsible for the long-term evolution. 
In order to obtain the estimation formula of the surface density at the vortex center, we use the fact that $\Sigma_\mathrm{v} r_\mathrm{v}^{1.5}$ remains almost constant. 
The vortencity conservation law explains the invariance of $\Sigma_\mathrm{v} r_\mathrm{v}^{1.5}$. 
It is likely that the estimation formula of the vortex migration speed is broadly applicable regardless of the formation mechanism of the vortex not only in 2D disks but also in 3D disks unless disks have very steep global slopes of the surface density or strong and wide pressure maxima. 
We also find that the fluid particles passing through the vortex region contribute to about 20\% of the total torque exerted on the vortex. 
In our interpretation, the remaining about 80\% of the total torque comes from the density waves. 
%check 2

Our calculations have been performed under a number of simplifying assumptions, but we consider we have captured some physical aspects of the RWI evolution and the RWI vortex. 
Our results provide a solid theoretical ground for quantitative interpretation of the observed lopsided structures in protoplanetary disks. 
Future studies considering other physical effects would allow us to make the models for the vortices that can be compared with observations. 
%check 2

\acknowledgments
Numerical computations were carried out using the Athena++ code on Cray XC40 at the 
Yukawa Institute Computer Facility, and on Cray XC30 at Center for Computational Astrophysics, National Astronomical Observatory of Japan. 
We would like to thank Samuel Richard for showing the snapshots of his calculations. 
We gratefully acknowledge Hideko Nomura, James Stone, Eugene Chiang, Jeffrey Fung, and Steve Lubow for their comments. 
We are also grateful to the referee who helped us improve the quality of the manuscript. 
This work was partially supported by Japan Society for the Promotion of Science (JSPS) KAKENHI Grant Numbers 15J01554 (T.O.),  26800106, 23103004, 15H02074, 17H01103 (T.M.), 16H05998, 16K13786 (K.T.). 
This research was also supported by The Ministry of Education,Culture,Sports,Science and Technology(MEXT) as ``Exploratory Challenge on Post-K computer" (Elucidation of the Birth of Exoplanets [Second Earth] and the Environmental Variations of Planets in the Solar System). 
\software{Athena++ code \citep[][https://princetonuniversity.github.io/athena/]{STW}}
%check 2

\appendix
\section{Analytic Steady Vortex Model}\label{sec:steady}
The steady solutions of vortices in shearing flow \citep{1981JPSJ..50..3517, 1987MNRAS.225..695G} are useful to understand the structure of the RWI vortices. 
In this section, we introduce the analytic steady vortex models. 
%check 2

We consider a  2D vortex orbiting a central star at angular velocity $\Omega_\mathrm{c}\ =\ \Omega(r_\mathrm{c})$ under the shearing box approximation, where $r_\mathrm{c}$ is the distance between the central star and the center of the vortex. 
The shearing box is a rotating Cartesian box centered at $r\ =\ r_\mathrm{c}$ with the angular velocity of $\Omega_\mathrm{c}$. 
We define $x\ =\ r-r_\mathrm{c}$ and $y\ =\ r_\mathrm{c} \varphi$ and neglect the terms arising from the cylindrical geometry. 
In this rotating frame, the equations of motion are 
\begin{eqnarray}
\diff{\bar{v}_x}{t}=\henb{\bar{v}_x}{t}+\bar{v}_x\henb{\bar{v}_x}{x}+\bar{v}_y\henb{\bar{v}_x}{y}&=&2\mathfrak{S}\Omega_\mathrm{c}^2x+2\Omega_\mathrm{c}\bar{v}_y-\henb{\Pi}{x}, \label{eq:ssx} \\
\diff{\bar{v}_y}{t}=\henb{\bar{v}_y}{t}+\bar{v}_x\henb{\bar{v}_y}{x}+v_y\henb{\bar{v}_y}{y}&=&-2\Omega_\mathrm{c}\bar{v}_x-\henb{\Pi}{y}, \label{eq:ssy}
\end{eqnarray}
where ${\bf \bar{v}}\ =\ (\bar{v}_x, \bar{v}_y)$ is the velocity field in the local shearing box. 
In equation (\ref{eq:ssx}), we have defined the mean shear without a vortex by $\mathfrak{S}\ \equiv\ -r\Omega_\mathrm{c}^{-1} [\partial \Omega\, /\, \partial r]_{r=r_\mathrm{c}}$, where $\mathfrak{S}\ =\ 1.5$ for a Keplerian disk. 
%check 2

In the steady vortex, the vorticity $\omega_z\ =\ (\mathrm{rot} {\bf \bar{v}})_z$ is assumed to be uniform. 
The fluid particles orbit the origin at constant angular velocity $\Omega_\mathrm{c}/\psi$ and the shape of the trajectories are elliptic, where $\psi$ is the turnover time of the vortex normalized by $2\pi/\Omega_\mathrm{c}$. 
We denote the semi-minor axis of the vortex by $b$, the semi-major axis of the vortex by $a$, and the vortex aspect ratio by $\chi \ \equiv \ a/b$. 
According to \citet{1981JPSJ..50..3517} (the equations (3.2) and (3.3)), the elliptic trajectories have to satisfy the following two conditions to be steady. 
First, the semi-minor axis should be aligned with $x-$ or $y-$axis. 
Otherwise, the elliptic trajectory would precess. 
When the semi-minor axis is aligned with the $x-$axis (radial direction), the velocity field is written as 
\begin{eqnarray}
\bar{v}_y &=& -\frac{\chi}{\psi}\Omega_\mathrm{c}x, \label{eq:kvy} \\
\bar{v}_x &=& \frac{1}{\chi \psi}\Omega_\mathrm{c}y. \label{eq:kvx} 
\end{eqnarray}
Defining the velocity field in the inertial frame by ${\bf v} \ =\ (v_x, v_y)$, 
we obtain from equations (\ref{eq:kvy}) and (\ref{eq:kvx}) 
\begin{eqnarray}
\henb{(v_y-v_K)}{x} &=& (\mathfrak{S}-\frac{\chi}{\psi})\Omega_\mathrm{c}, \label{eq:kvye} \\
\henb{v_x}{y} &=& \frac{1}{\chi \psi}\Omega_\mathrm{c}, \label{eq:kvxe} 
\end{eqnarray}
where $v_\mathrm{K}~=~r_\mathrm{c}\Omega_\mathrm{c}+(1-\mathfrak{S})\Omega_\mathrm{c}x$ is the Keplerian rotation velocity. 
Second, the vortex aspect ratio $\chi$ and the turnover time $\psi$ should satisfy 
\begin{equation}
(\chi -1) = \mathfrak{S} \psi, \label{eq:cpre}
\end{equation}
for the invariance of the vortex aspect ratio. 
The analytic vortex solution which satisfies these conditions is called the Kida solution. 
%check 2

\citet{1987MNRAS.225..695G} derived another relation between $\chi$ and $\psi$. 
They assume the velocity field satisfying equations (\ref{eq:kvy})--(\ref{eq:kvx}) and stationary compressible flow in the shearing box. 
Substituting equations (\ref{eq:kvy}) and (\ref{eq:kvx}) into equations (\ref{eq:ssx}) and (\ref{eq:ssy}) and assuming $\partial\, /\, \partial t\ =\ 0$ (steady state), the pressure function $\Pi$ should satisfy 
\begin{eqnarray}
\henb{^2\Pi}{x^2} &=& (\frac{1}{\psi^2}-2\frac{\chi}{\psi}+2\mathfrak{S})\Omega_\mathrm{c}^2, \label{eq:pfxx} \\
\henb{^2\Pi}{y^2} &=& (\frac{1}{\psi^2}-2\frac{1}{\chi \psi})\Omega_\mathrm{c}^2. \label{eq:pfyy} 
\end{eqnarray}
Substituting equations (\ref{eq:pfxx}) and (\ref{eq:pfyy}) into the continuity equation, another relation between $\chi$ and $\psi$, 
\begin{equation}
(\chi^2-1)=2\mathfrak{S}\psi^2, \label{eq:gore}
\end{equation}
is obtained \citep{1987MNRAS.225..695G}. 
The analytic vortex solution which satisfies rather equation (\ref{eq:gore}) than equation (\ref{eq:cpre}) is called the GNG solution. 
%check 2

The Kida solution (equation (\ref{eq:cpre})) is not compatible with the GNG solution (equation (\ref{eq:gore})) except for  
\begin{eqnarray}
\chi&=&\frac{2+\mathfrak{S}}{2-\mathfrak{S}} = 7, \label{eq:chg} \\
\psi&=&\frac{2}{2-\mathfrak{S}} = 4. \label{eq:tog} 
\end{eqnarray}
Here, we have assumed $\mathfrak{S}\ =\ 1.5$ (Keplerian shear) in the last equalities. 
Note that the steady solution gives us the gradients of the velocity field and the convexities of the pressure function around the vortex center, but does not give any information about the size or the surface density in the vortices. 
%check 2

\section{Diagnostics of the Axisymmetric Disk Profiles}
We introduce two criteria to assess axisymmetric disks. 
In this section, we assume the axisymmetric disks. 

\subsection{Rayleigh's Condition}
When there is a radius at which  
\begin{equation}
\kappa^2 (r) \equiv \frac{1}{r^3}\diff{[rv_{\varphi}(r)]}{r}<0, 
\end{equation}
is satisfied, where $\kappa$ is the epicyclic frequency, the gas distribution is unstable to the rotational instability, which is an axisymmetric hydrodynamical instability in differentially rotating disks \citep{1960PNAS...46..253C}. 
This is known as Rayleigh's criterion. 
We use the term ``Rayleigh's condition'' when there is a radius where Rayleigh's stability criterion is violated. 
%check 2

In the case that the initial surface density distribution is given by a Gaussian bump, the Rayleigh's condition can be regarded as a sufficient condition for the onset of the RWI. 
We note that the Rayleigh's condition is not always the sufficient condition for the RWI in general (see paper I). 
When $\mathcal{A}_0$ is large or $\Delta w_0$ is small, the disk is unstable to the rotational instability. 
In this paper, we set parameters so that the Rayleigh's condition is not violated. 
In other words, the Rayleigh's condition gives the upper limit of $\mathcal{A}_0$ for each $h$ and $\Delta w_0$ in our parameter space. 
%check 2

\subsection{The Critical Condition for the RWI}
For a barotropic flow, the RWI requires that the vortensity, $q (r)\ \equiv \ \kappa^2/(2\Sigma \Omega)$, has a local minimum \citep[][see also Paper I]{1999ApJ...513..805L}. 
This is a necessary condition but is not sufficient. 
We found the way to explore parameters where the disk is marginally stable to the RWI (see Section 5.2 in paper I). 
In this paper, we use this condition to determine the lower limit of $\mathcal{A}_0$ for each $h$ and $\Delta w_0$ in our parameter space. 
%check 2

In paper I, we also derived the necessary and sufficient conditions for the onset of the RWI with the azimuthal mode $m$ in a semi-analytic form as 
\begin{equation}
\eta_m\equiv \int^{r_\mathrm{OR}}_{r_\mathrm{IR}}\sqrt{-D_{\mathrm{MS},m}}\, \mathrm{d}r \gtrsim \eta_\mathrm{c} \approx \frac{\pi}{2\sqrt{2}}, \label{eq:bcod}
\end{equation}
where 
\begin{eqnarray}
D_{\mathrm{MS},m}(r)&\equiv& \frac{1}{2}\diff{B_{\mathrm{MS},m}}{r}+\frac{B_{\mathrm{MS},m}^2}{4}-C_{\mathrm{MS},m}, \label{eq:Dmsdef} \\
B_{\mathrm{MS},m}(r)&\equiv& \diff{\, \mathrm{ln}\, (rF_{\mathrm{MS},m}/\Omega)}{r}\\
C_{\mathrm{MS},m}(r)&\equiv& -\frac{m^2}{r^2}-\frac{\kappa^2-m^2(\Omega_q-\Omega)^2}{c_{\mathrm{s}}^2}-\frac{2}{r}\frac{\Omega}{(\Omega_q-\Omega)}\diff{\, \mathrm{ln}\, (F_{\mathrm{MS},m})}{r},\\
F_{\mathrm{MS},m}(r)&\equiv&\frac{\Sigma \Omega}{\kappa^2-m^2(\Omega_q-\Omega)^2}, \label{eq:Fmsdef}
\end{eqnarray}
$\Omega_q$ is the angular velocity at the local minimum of the vortensity $q$, and $c_{\mathrm{s}}\ \equiv \ \sqrt{\Gamma P/\Sigma}$ is the adiabatic sound speed, and $r_\mathrm{IR}$ and $r_\mathrm{OR}$ are  the radii where $D_{\mathrm{MS},m}$ vanishes. 
Since equation (\ref{eq:bcod}) is derived using the Sommerfeld--Wilson quantization condition \citep{Wilson1915, Sommerfeld1916}, $\eta_\mathrm{c}$ is equal to $\pi/(2\sqrt{2})$ only when the azimuthal mode $m$ is large or the shape of the $D_{\mathrm{MS},m}$ potential well is perfectly parabolic. 
We expect $\eta_\mathrm{c}\ >\ \pi/(2\sqrt{2})$ for a small $m$ mode and shallower $D_{\mathrm{MS},m}$ potential well, and $\eta_\mathrm{c}\ <\ \pi/(2\sqrt{2})$ for a small $m$ mode and steeper $D_\mathrm{MS}$ potential well from the knowledge of quantum mechanics. 
%check 2

We also use equation (\ref{eq:bcod}) to investigate the stability of the axisymmetric components against the RWI during the RWI evolution in Section 3.2 and Section 3.3. 
At that time, we calculate the values defined by equations (\ref{eq:Dmsdef})--(\ref{eq:Fmsdef}) using the radial profiles of the azimuthally averaged surface density and the azimuthally averaged rotation velocity instead of those in the initial profiles. 
%check 2

\section{Supplementary Data}
\subsection{The Linear Growth Rate of the Initial Conditions of Each Azimuthal Mode}
In Table \ref{tbl:init_sub1}, we show the most unstable azimuthal mode, $m_\ast$, and the linear growth rates for $1\ \leq \ m \ \leq \ 10$ modes, $\gamma_m$, calculated by the same method as in paper I. 
For all the linear stability analyses, we adopt the inner radius $r_\mathrm{in} \ =\ 0.3$ and the outer radius $r_\mathrm{out}\ =\ 2.5$. 
We set the radial grid number $N$ as $N\ =\ 1024$ for $h\ =\ 0.1,\ 0.15,\ 0.2$, and $N\ =\ 2048$ for $h\ =\ 0.05$.  
%check 2

\begin{table*}
\begin{center}
\caption{The most unstable azimuthal mode and the linear growth rates of each mode. \label{tbl:init_sub1}}
\renewcommand{\arraystretch}{0.85}
\begin{tabular}{lc|cccccccccc}
\tableline \tableline
&&\multicolumn{10}{c}{linear growth rate for each azimuthal mode $(\gamma_m/\Omega_\mathrm{n})$} \\ \cline{3-12}
Name &$m_\ast$&1&2&3&4&5&6&7&8&9&10 \\
\hline 
\hline 
h10w1g1&9&0.053&0.100&0.141&0.172&0.195&0.210&0.221&0.226&{\bf 0.227}&0.225\\
h10w1g2&8&0.053&0.095&0.131&0.158&0.177&0.190&0.197&{\bf 0.200}&0.199&0.194\\
h10w1g3&7&0.042&0.079&0.108&0.128&0.141&0.148&{\bf 0.150}&0.148&0.141&0.131\\
h10w1g4&6&0.033&0.061&0.081&0.094&0.0996&{\bf 0.1000}&0.096&0.087&0.074&0.054\\
h10w1g5&4&0.021&0.038&0.047&{\bf 0.050}&0.047&0.039&0.026&0.008&--&--\\
h10w2g1&6&0.074&0.140&0.189&0.220&0.238&{\bf 0.242}&0.237&0.222&0.199&0.168\\
h10w2g2&5&0.067&0.123&0.165&0.190&{\bf 0.201}&0.200&0.190&0.170&0.142&0.106\\
h10w2g3&5&0.054&0.100&0.131&0.147&{\bf 0.150}&0.142&0.124&0.098&0.062&0.012\\
h10w2g4&4&0.042&0.075&0.095&{\bf 0.100}&0.094&0.077&0.052&0.017&--&--\\
h10w2g5&3&0.026&0.044&{\bf 0.050}&0.043&0.025&0.002&--&--&--&--\\
h10w3g1&4&0.103&0.184&0.232&{\bf 0.246}&0.230&0.188&0.122&0.040&--&--\\
h10w3g2&4&0.089&0.158&0.195&{\bf 0.200}&0.177&0.130&0.059&0.001&--&--\\
h10w3g3&3&0.072&0.126&{\bf 0.150}&0.145&0.113&0.059&0.002&--&--&--\\
h10w3g4&3&0.054&0.090&{\bf 0.100}&0.089&0.041&--&--&--&--&--\\
h10w3g5&2&0.033&{\bf 0.050}&0.042&0.009&--&--&--&--&--&--\\
h10w4g1&3&0.140&0.224&{\bf 0.227}&0.149&0.005&--&--&--&--&--\\
h10w4g2&2&0.126&{\bf 0.200}&0.195&0.115&0.002&--&--&--&--&--\\
h10w4g3&2&0.098&{\bf 0.150}&0.131&0.044&--&--&--&--&--&--\\
h10w4g4&2&0.070&{\bf 0.100}&0.065&--&--&--&--&--&--&--\\
h10w4g5&2&0.043&{\bf 0.050}&0.002&--&--&--&--&--&--&--\\
h10w5g1&2&0.174&{\bf 0.191}&0.002&--&--&--&--&--&--&--\\ 
h10w5g3&2&0.146&{\bf 0.150}&0.001&--&--&--&--&--&--&--\\ 
h10w5g4&1&{\bf 0.100}&0.080&--&--&--&--&--&--&--&--\\ 
h10w5g5&1&{\bf 0.050}&0.00&--&--&--&--&--&--&--&--\\ \hline
h20w1g1&8&0.064&0.118&0.153&0.177&0.191&0.201&0.206&{\bf 0.209}&0.208&0.206\\
h20w1g4&4&0.044&0.076&0.093&{\bf 0.100}&0.099&0.096&0.086&0.076&0.061&0.048\\
h20w2g1&5&0.093&0.162&0.201&0.220&{\bf 0.224}&0.220&0.205&0.186&0.159&0.132\\
h20w2g4&3&0.056&0.091&{\bf 0.100}&0.093&0.072&0.046&0.005&--&--&--\\
h20w3g1&3&0.126&0.207&{\bf 0.237}&0.229&0.195&0.140&0.063&0.001&--&--\\
h20w3g4&2&0.069&{\bf 0.100}&0.087&0.045&--&--&--&--&--&--\\
h20w4g1&2&0.163&{\bf 0.237}&0.213&0.118&--&--&--&--&--&--\\
h20w4g4&2&0.087&{\bf 0.100}&0.030&--&--&--&--&--&--&--\\
h20w5g1&2&0.191&{\bf 0.192}&0.020&--&--&--&--&--&--&--\\ 
h20w5g4&1&0.100&{\bf 0.049}&--&--&--&--&--&--&--&--\\\hline
h15w1g1&8&0.059&0.112&0.150&0.177&0.195&0.207&0.213&{\bf 0.216}&0.215&0.212\\
h15w1g4&5&0.039&0.070&0.089&0.098&{\bf 0.100}&0.097&0.089&0.079&0.065&0.049\\
h15w2g1&5&0.086&0.155&0.199&0.224&{\bf 0.233}&0.231&0.219&0.200&0.175&0.144\\
h15w2g4&3&0.050&0.085&{\bf 0.100}&0.098&0.084&0.060&0.028&--&--&--\\
h15w3g1&4&0.117&0.200&0.238&{\bf 0.240}&0.213&0.162&0.092&0.007&--&--\\
h15w3g4&2&0.064&{\bf 0.100}&0.099&0.084&0.060&0.028&--&--&--&--\\
h15w4g1&2&0.153&{\bf 0.233}&0.222&0.136&0.003&--&--&--&--&--\\
h15w4g4&2&0.079&{\bf 0.100}&0.048&--&--&--&--&--&--&--\\
h15w5g1&2&0.183&{\bf 0.193}&0.025&--&--&--&--&--&--&--\\
h15w5g4&1&{\bf 0.100}&0.066&--&--&--&--&--&--&--&--\\\hline
h05w1g1&11\tablenotemark{$\dagger$}&0.048&0.078&0.116&0.149&0.176&0.199&0.216&0.229&0.238&0.243\\
h05w1g4&7&0.023&0.044&0.063&0.079&0.090&0.097&{\bf 0.100}&0.099&0.095&0.088\\
h05w2g1&7&0.057&0.112&0.160&0.198&0.225&0.242&{\bf 0.248}&0.244&0.232&0.211\\
h05w2g4&5&0.030&0.058&0.080&0.094&{\bf 0.100}&0.098&0.087&0067&0.039&--\\
h05w3g1&4&0.085&0.159&0.213&{\bf 0.240}&0.237&0.206&0.147&0.064&0.011&--\\
h05w3g4&4&0.044&0.079&{\bf 0.100}&{\bf 0.240}&0.237&0.206&0.147&0.064&0.010&--\\
h05w4g1&3&0.128&0.214&{\bf 0.225}&0.151&0.001&--&--&--&--&--\\
h05w4g4&2&0.064&{\bf 0.100}&0.082&0.001&--&--&--&--&--&--\\
h05w5g1&2&0.168&{\bf 0.188}&0.001&--&--&--&--&--&--&--\\ 
h05w5g4&1&{\bf 0.100}&0088&--&--&--&--&--&--&--&--\\ 
\tableline \tableline
\\
\multicolumn{12}{p{.8\textwidth}}{NOTE. Name: the name of the model. $m_\ast$: the most unstable azimuthal mode. $\gamma_m/\Omega_\mathrm{n}$: the linear growth rate against the RWI for each azimuthal mode $m$. }
\end{tabular}
\end{center}
\tablenotetext{\dagger}{The largest linear growth rate $(\gamma_\ast/\Omega_\mathrm{n})$ is $0.244$ in the "h05w1g1" model.}
\end{table*}

\subsection{Table of the Results}
We show the values of $\tau_2$, $\tau_1$, and $\tau_2-\tau_1$ in Table \ref{tbl:tau}. 
For the ``h10w5g3''--``h10w5g5'', ``h20w5g4'', ``h15w5g4'', and ``h05w5g4'' models, we are not able to measure $\tau_2$ and $\tau_1$ with visual inspection due to low $m_\ast$. 
In those cases, we set $\tau_2$ to no data and $\tau_1$ to the orbit number at the saturation. 
From the long-term calculations, we also show the values of $\tau_\chi$ in Table \ref{tbl:tau}. 
For the ``h10w2g1'', ``h10w3g1'', ``h10w4g1'', ``h10w4g2'', ``h10w5g1'', ``h10w5g2'', ``h20w1g1'', ``h20w2g1'', ``h20w3g1'', ``h20w4g1'', ``h20w5g1'', ``h15w1g1'', ``h15w2g1'', ``h15w3g1'', ``h15w4g1'', ``h15w5g1'', ``h05w1g1'', ``h05w2g1'', ``h05w4g1'', and ``h05w5g1'' models, we are not able to measure $\tau_\chi$ due to fast vortex migration. 
In those cases, we set $\tau_\chi$ to no data. 
%check 2

\begin{table*}
\caption{Values of $\tau_2,\ \tau_1,\ \tau_1 - \tau_2$, and $\tau_\chi$. \label{tbl:tau}}
\begin{center}
\renewcommand{\arraystretch}{1.0}
\begin{tabular}{ll}
\begin{tabular}[t]{lcccc}
\tableline \tableline
Name & $\tau_2$ & $\tau_1$ &$\tau_1~-~\tau_2$&$\tau_\chi$\\
\hline 
\hline 
h10w1g1&23&45&22&5.2E3\\
h10w1g2&33&57&24&3.6E3\\
h10w1g3&36&96&60&3.7E3\\
h10w1g4&52&66&14&3.0E3\\
h10w1g5&54&98&44&1.3E3\\
h10w2g1&24&84&60&--\\
h10w2g2&21&46&25&2.4E4\\
h10w2g3&21&111&90&3.6E3\\
h10w2g4&28&49&21&3.1E3\\
h10w2g5&41&109&68&2.8E3\\
h10w3g1&18&20&2&--\\
h10w3g2&20&45&28&5.8E3\\
h10w3g3&22&42&20&8.9E3\\
h10w3g4&25&35&10&8.0E3\\
h10w3g5&47&72&25&5.0E3\\
h10w4g1&10&17&7&--\\
h10w4g2&12&20&8&--\\
h10w4g3&14&28&14&7.3E3\\
h10w4g4&20&36&16&1.4E4\\
h10w4g5&36&45&9&1.7E4\\
h10w5g1&12&13&1&--\\
h10w5g3&--&15&--&--\\
h10w5g4&--&22&--&3.3E3\\
h10w5g5&--&52&--&5.0E3\\\hline
h20w1g1&24&25&1&--\\
h20w1g4&30&59&29&2.2E3\\
h20w2g1&14&27&13&--\\\hline
\tableline \tableline
\end{tabular}
&
\begin{tabular}[t]{lcccc}
\tableline \tableline
Name & $\tau_2$ & $\tau_1$ &$\tau_1~-~\tau_2$&$\tau_\chi$\\
\hline 
\hline 
h20w2g4&21&37&16&1.3E3\\
h20w3g1&12&15&3&--\\
h20w3g4&29&42&13&1.4E3\\
h20w4g1&8&16&8&--\\
h20w4g4&19&26&7&2.7E3\\
h20w5g1&11&12&1&--\\
h20w5g4&--&23&--&2.0E3\\\hline
h15w1g1&37&62&25&--\\
h15w1g4&30&82&52&3.3E3\\
h15w2g1&16&27&11&--\\
h15w2g4&25&32&7&1.1E3\\
h15w3g1&14&21&7&--\\
h15w3g4&19&35&16&2.0E3\\
h15w4g1&8&17&9&--\\
h15w4g4&19&29&10&3.0E3\\
h15w5g1&11&13&2&--\\
h15w5g4&--&23&--&2.0E3\\\hline
h05w1g1&69&177&108&--\\
h05w1g4&57&195&138&1.4E4\\
h05w2g1&29&199&170&--\\
h05w2g4&36&182&146&2.1E4\\
h05w3g1&16&45&29&9.2E3\\
h05w3g4&26&49&23&6.8E3\\
h05w4g1&10&21&11&--\\
h05w4g4&19&46&27&5.2E3\\
h05w5g1&13&15&2&--\\
h05w5g4&--&24&--&2.1E4\\\hline
\tableline \tableline
\end{tabular}
\\
\\
\multicolumn{2}{p{.8\textwidth}}{NOTE. Name: the name of the model. $\tau_2$: the orbit when the number of the vortices becomes two. $\tau_1$: the orbit when the final vortex merger occurs. $(\tau_1-\tau_2)$: the duration of the two vortices regime. $\tau_\chi$: the decreasing time of $\chi_2$ in the long-term calculations. }
\end{tabular}
\end{center}
\end{table*}

\section{Numerical Test}

In order to check the convergence of our numerical calculations, we have additionally performed a high-resolution calculation and a low-resolution calculation of the ''h10w3g1'' model. 
The high-resolution calculation has twice as many cells as the fiducial calculation does in each direction. 
In contrast, the low-resolution calculation has half cells compared to the fiducial calculation in each direction. 
We also perform wide-domain calculations in order to check the effects of the inner boundary on the RWI vortices.  
In the wide-domain calculations, the resolution is almost the same as that of the fiducial calculation, but the radius of the inner boundary is set at $r_\mathrm{in}\ =\ 0.2r_\mathrm{n}$, $0.1r_\mathrm{n}$, and $0.03r_\mathrm{n}$ instead of $0.3r_\mathrm{n}$, respectively. 
%check 2

\begin{figure}
\epsscale{0.6}
\plotone{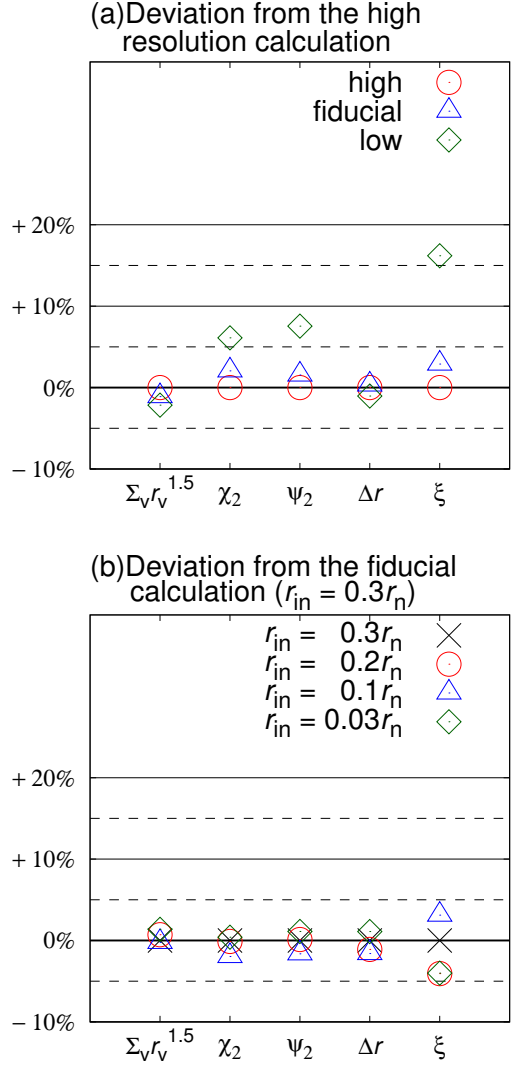}
\vspace{-1.35cm}
\caption{Panel (a) shows the deviations of the parameters ($\Sigma_\mathrm{v}r_\mathrm{v}^{1.5}$, $\chi_2$, $\psi_2$, $\delta r$, and $\xi$) from the high-resolution calculation in the high-resolution calculation (the red circles), the fiducial calculation (the blue triangles), and the low-resolution calculation (the green diamonds). Panel (b) shows the deviations of the parameters from the fiducial calculations in the wide-range calculations ($r\mathrm{in}\ =\ 0.3r_\mathrm{n}$: the black cross point, $r\mathrm{in}\ =\ 0.2r_\mathrm{n}$: the red circles, $r\mathrm{in}\ =\ 0.1r_\mathrm{n}$: the blue triangles, and $r\mathrm{in}\ =\ 0.03r_\mathrm{n}$: the green diamonds).}
\label{fig:res1}
\end{figure}

In all the calculations, the RWI vortices are formed at almost the same time ($\tau \ \approx \ 20$). 
We measure 5 parameters ($\Sigma_\mathrm{v}r_\mathrm{v}^{1.5}$, $\chi_2$, $\psi_2$, $\Delta$r, and $\xi$) that characterize the vortex and its migration and are almost constant in each additional calculation due to the fast vortex migration. 
Figure \ref{fig:res1} shows the deviation of the parameters from those of the high-resolution calculation. 
Since the RWI vortex in the fiducial calculation shows the same values of parameters as those in the high-resolution calculation within 5\% (see Panel (a) of Figure \ref{fig:res1}), we conclude that the fiducial calculation has sufficiently high-resolution and the results are converged. 
Panel (b) of Figure \ref{fig:res1} shows that the values in the wide-domain calculations are also equivalent to those in the fiducial calculation within 5\%. 
This indicates that the inner boundary does not have significant effects on the RWI vortex. 
%check 2

\end{document}